 \documentclass[preprint2]{aastex}

\slugcomment{Not to appear in Nonlearned J., 45.}

\shorttitle{Collapsed Cores in Globular Clusters}
\shortauthors{Djorgovski et al.}

\begin{document}


\title{Statistically Stable Estimates  of Variance in Radioastronomical Observations as Tools for  RFI Mitigation}


\author{P. A. Fridman}
\affil{ASTRON, Dwingeloo, Postbus 2, 7990AA, The Netherlands}
\email{fridman@astron.nl}



\begin{abstract}
A selection of statistically stable (robust)  algorithms for data
variance calculating  has been made. Their properties have been
analyzed via computer simulation. These algorithms would be useful
if adopted in  radio astronomy observations in the presence of
strong sporadic radio frequency interference (RFI). Several
observational results have been presented here to demonstrate the
effectiveness of these algorithms in RFI mitigation.
\end{abstract}


\keywords{methods: miscellaneous --- methods: statistical}


\section{Introduction}

    Cosmical radio emissions received by radio telescopes are  noise-like signals  which are characterized  by a {\it normal}  (Gaussian) probability distribution function ${\cal N}(0,\sigma_{sig})$, i.e., with zero mean and  variance $\sigma_{sig}^2$. Background radio emission and radio receivers also produce   normal  noise  ${\cal N}(0,\sigma_{sys})$. Given a large number of n samples $x_i, i=0..n$ of  normally distributed data with  zero mean, a classical statistical  procedure for the estimation of  variance $\sigma^2$ is
\begin{equation}
\widehat{\sigma^2_{n}}=\frac{1}{n}\sum_{i=1}^{n}x_{i}^2
\end{equation}
The time-honoured estimate (1) may be obtained by using the Maximum
Likelihood method \citep{warden69} and is a {\it minimum variance
unbiased estimate}. Traditional radiometrical techniques implement
this procedure in analogue or digital form in conventional
radiometers: the estimate of $\widehat{\sigma_{sig}^2
+\sigma_{sys}^2}$ is obtained after   square-law detection and
low-pass filtering (averaging) \citep{rohlfs06}. This statistical
inference is based  upon an essential assumption about the
probability distribution of the data  ``signal noise + systen
noise'': the distribution must be normal. As in many cases in
applied statistics this assumption is not always valid for radio
astronomy. Most of the data   may correspond to the model  with an
assumed  normal distribution, but there are also a number of {\it
outliers}, atypical data which  stand out from the bulk of the data.
As a result there is an  ``approximately''  normal distribution that
gives rise to  outliers, i.e., the distribution has a normal shape
in the central region  but  has tails that are heavier  than those
of a normal distribution. If such approximate normality were to
hold, the results of using a normal distribution theory will
nevertheless not hold approximately. In the presence of heavy tails
the estimate (1) based on the maximum likelihood principle  is no
longer the best and may have unacceptably low statistical efficiency
(large variance) and  very large bias. These phenomena in applied
statistics have been well known from the time of Gauss, Newcomb and
Eddington.

Radio frequency interference (RFI) creates a situation in radio
astronomy where such outliers arise \citep{fridman01, kest07}.
Industry, ground and satellite communications, ground and airborne
radar, power lines, radio and TV stations, etc. produce all kinds of
additional noise which penetrate into  extremely sensitive radio
astronomy receivers. Radio observatory  computers are themselves
often the sources of RFI. Many authors demonstrated the
effectiveness of different RFI mitigation methods
 \citep{elli03, baan04, fisher05, mitch05, poul05, Zhang05, kest07} in observations with single dishes and radio interferometers.  Adaptive noise cancellation (ANC)  with reference antenna was described in \citep{barn98}. Post detector and post correlation processing using  reference antennas was suggested in \citep{briggs01}. Higher order statistics (HOS) analysis was proposed in \citep{frid01a}.
Principal results were achieved with  blanking and  reference
antennas.  The high expectations of new projects (LOFAR, ATA, SKA)
are based on the implementation of spatial processing - nulling in
the direction of RFI.  But very often RFI do not obey the model
assumptions which guarantee the success of this kind of processing.
They are variable, sporadic and are not  fully coherent at the sites
of  multi-element systems \citep{thom03, jeffs05}.

In this paper  a new view on blanking is proposed based on
statistical methods granting stable estimates of variance in the
presence of outliers. These methods would be useful in radio
astronomy RFI mitigation, including  {\it real-time} processing. The
aim of this paper is to  compile  a selection of  methods of stable
variance estimates suitable for RFI mitigation   and to study their
properties via computer simulation. Some examples obtained during
real observations are also given.

The typical appearances of RFI  is of strong impulse-like bursts  in
both temporal and spectral domains.  They behave as outliers in the
data with normal distribution and  the   data  received may be
characterized by the {\it contaminated} normal distribution
\begin{equation}
F(x)=(1-\epsilon){\cal N}(0,\sigma_{sys})+\epsilon F_{RFI}(x),
\end{equation}

where ${\cal N}(0,\sigma_{sys})$ is the ``clean'' probability
distribution, $F_{RFI}(x)$ is the usually unknown distribution of
RFI, $\epsilon$ characterizes the fraction of  $F_{RFI}(x)$ in the
total $F(x), 0<\epsilon<1$. The theory of robustness   pioneered by
\citep{tukey60, huber64, hampel71} gives recommendations on
obtaining  statistically stable estimates of parameters in the
situation similar to (2).   The estimate is called statistically
stable (robust) if slight changes in distribution have a relatively
small effect on its value, or, in other words, robustness means
insensitivity of a statitical procedure  to small deviations from
assumptions about the normality of data's  distribution.

Several approaches to developing robust estimates exist:
$M$-estimates based on a generalized maximum likelihood \citep
{huber04}, $L$-estimates based on order statistics \citep{david03},
$R$-estimates based on  ranks \citep{hett84} and estimates based on
nonparametrical statistics \citep{sheskin00}.

\section{Stable estimates of variance}
A selection of stable (robust) estimates of variance will be
described in this section. There are different ways of judging the
stability of an estimate. Given a distribution $F(x)$ and a
population $X$ representing  this distribution, an estimate $T(X,F)$
is the functional depending on $F(x)$. If this estimate is
relatively unaffected by small changes in $F$, i. e., the functional
is continuous  with respect to $F$, then the estimate is said to
have  {\it qualitative robustness}.  For a distribution $F(x)$
contaminated by a distribution $G(x)$ with probability $\epsilon$
 \begin{equation}
F_{\epsilon}(x)=(1-\epsilon)F(x)+\epsilon G(x)
\end{equation}
the  {\it influence  function} \citep{hampel86}  $IF(T,F)$ is widely
used in theoretical analysis
\begin{equation}
IF(x)=\lim_{\epsilon \rightarrow 0}\frac{T(F_{x,\epsilon
})-T(F_{x})}{\epsilon }
\end{equation}
The $IF(T,F)$ is the relative influence of x on $T(F)$  having the
value $x$ with probability $\epsilon$. The estimate $T(F)$ is said
to have {\it infinitesimal robustness}  if $IF(x)$  is bounded. The
$IF$  of classical  both arithmetical mean and variance are not
bounded. A simple finite-sample version of  $IF$ called {\it
empirical influence function} (or sensitivity function) exists. It
is constructed in the following way. Suppose there is an estimate
$T_{n-1}$ of a sample $(x_{1},...,x_{n-1})$. A new $n-th$ sample $x$
is added and $T_{n}(x_{1},...,x_{n-1},x)$  is considered as a
function of x.  For the estimate (1) the sensitivity function is
\begin{equation}
IF(x)=\frac{x^{2}-\widehat{\sigma ^{2}_{n}}}{n+1}
\end{equation}
Another useful characteristic of robustness is the {\it breakdown
point}: the minimum value  of $\epsilon$ for which an estimate goes
to infinity as  $x$ grows. The breakdown point of conventional mean
and variance is 0. The breakdown point of median is 0.5, i.e., up to
50\% of data may be outliers and they do not alter the correct value
of the median.

Taking into consideration our practical goal to use  estimates of
variance in observations,  two  measures of stability will be
exploited for the  following  statistical procedures.

1. The relative empirical influence function with  the following
modification: for a sample length $N$ the number of additional
outliers with an amplitude $X_{RFI}$ will be $M=[\epsilon N]$, where
$[q]$ means  the greatest integer part of $q$. The estimate
calculated for a contaminated sample is compared with the estimate
correspondimg to a ``clean'' sample:
 \begin{equation}
REIF(\epsilon x)=\frac{T_{N+M}(F_{x,\epsilon
})-T_{N+M}(F_{x,\epsilon =0})}{T_{N+M}(F_{x,\epsilon =0})}.
\end{equation}
The relative empirical influence function will be calculated during
computer simulation for the estimates described below.

2. Stability against outliers is achieved  at the expense  of the
effectiveness of an estimate. In the absence of outliers the
standard deviation of a robustly estimated variance is, as a rule,
larger  than that of a simple estimate (1). To characterize this
loss the parameter $LOSS$ will be used:
\begin{eqnarray}
LOSS=SNR_{T_{N}}/SNR0,\\
SNR_{T_{N}}=\frac{\widehat{T_{N}(\sigma +\Delta \sigma )}-\widehat{T_{N}(\sigma )}}{rms[\widehat{T_{N}(\sigma )}]},\nonumber\\
SNR0=\frac{\widehat{T0_{N}(\sigma +\Delta \sigma
)}-\widehat{T0_{N}(\sigma )}}{rms[\widehat{T0_{N}(\sigma
)}]}.\nonumber
\end{eqnarray}

For  a small increase of $\Delta \sigma << \sigma$ two
signal-to-noise ratios $SNR$ are compared: $SNR_{T_{N}}$ of a given
estimate  of variance $\widehat{T_{N}(\sigma )}$ and $SNR0$ of the
estimate (1)  with the best potential effectiveness.

The relative empirical influence function and loss are calculated
via computer simulations ($\sigma=1, n=10^{5}$)  for the estimates
presented below.

\subsection{ Variance of the trimmed data}
Let  $x_{1},...., x_{n}$ be a random sample and let $x_{(1)} \leq
x_{(2)}\leq...\leq x_{(n)}$ be the observations sorted in ascending
order. The $ith$ largest value $x_{(i)}$ is called the $ith$ {\it
order statistic}. Let $\gamma $ denote  the chosen amount of
trimming, $0 \leq \gamma  \leq 0.5$ and $k=[\gamma  n]$.
 The sample trimmed variance is computed by removing the $k$  largest and  $k$ smallest data
and using the values that remain:
\begin{eqnarray}
T_{1}=\frac{K_{trim}}{N-2k}\sum_{n=k}^{N-k}(x_{n}-\widehat{\mu }_{trim})^{2}\\
\widehat{\mu }_{trim}=\frac{1}{N-2k}\sum_{n=k}^{N-k}x_{n}, \nonumber
\end{eqnarray}
where $\mu_{trim}$ is the sample mean of the trimmed data. Trimming
lessens the variance of  data and the coefficient $K_{trim}$ makes
$T_{1}$ the consistent estimator for  data with  normal
distribution. Table  1 gives the values of $K_{trim}$  for different
$\gamma$.

\begin{table}[h]
\caption{Consistency factor and $LOSS$ as functions of $\gamma$
for trimming}
\begin{tabular}{llllll}
\tableline \tableline
$\gamma$ & 0.005 & 0.01 & 0.025 & 0.05 & 0.1 \\
$K_{trim}$ & 1.085 & 1.147 & 1.32 & 1.6 & 2.28 \\
$LOSS$ & 0.98 & 0.95 & 0.90 & 0.86 & 0.83 \\
\tableline
\end{tabular}
\end{table}
The third row in this table gives the values of $LOSS$ for different $\gamma$.\\
Fig. 1 shows the relative empirical influence functions for
untrimmed data (Fig. 1a) and for the variances computed for trimmed
data with $\gamma=0.05$ (Fig. 1b). The parameter $\epsilon$
(``eps'') indicates  the percentage of outliers in the total volume
of data.

\subsection{Winsorized sample variance}
A sample $x_{1},...., x_{n}$ is sorted in ascending order.  For the
chosen $0 \leq \gamma  \leq 0.5$ and $k=[\gamma  n]$ winsorization
of the sorted data consists of setting
\begin{equation}
W_{i}=\left\{ \begin{array}{lcl}x_{(k+1)}, & if & x_{(i)}\leq x_{(k+1)}\\
                                x_{(i)},   & if  & x_{(k+1)}<x_{(i)}<x_{(n-k)}\\
                    x_{(n-k)}, & if & x_{(i)}\geq x_{(n-k})\\

  \end{array} \right.
\end{equation}

The winsorized sample mean is $\widehat{
\mu_{w}}=\frac{1}{n}\sum_{i=1}^{n}W_{i}$ and the winsorized sample
variance is
\begin{equation}
T_{2}=\frac{1}{n-1}\sum_{i=1}^{n}(W_{i}-\widehat{\mu _{w}})^{2}
\end{equation}

Table 2 gives values of the consistency factor $K_{winsor}$  for
different $\gamma$.

\begin{table}[h]
\caption{Consistency factor and $LOSS$ as functions of $\gamma$
for winsorization}
\begin{tabular}{llllll}
\tableline \tableline
$\gamma$ & 0.005 & 0.01 & 0.025 & 0.05 & 0.1 \\
$K_{winsor}$ & 1.019 & 1.041 & 1.103 & 1.227 & 1.54 \\
$LOSS$ & 0.99 & 0.98 & 0.96 & 0.88 & 0.82 \\
\tableline
\end{tabular}
\end{table}
The third row of this table  gives the values of $LOSS$ for different $\gamma$.\\
Fig. 1c shows the relative empirical influence functions  for the
variances computed for winsorized data with $\gamma=0.1$. The
parameter $\epsilon$ (``eps'') indicates the percentage of outliers
in the total volume  of data.

\subsection{ Median absolute deviation}

This estimate for  sorted data $x_{(1)} \leq x_{(2)}\leq...\leq
x_{(n)}$ is defined by
\begin{equation}
T_{3}=1.483\times med_{1\leq i\leq n}\{\left|
x_{i}-med(x_{i})\right| \},
\end{equation}
where
\begin{displaymath}
\left. \begin{array}{lll}
med= & 0.5(x_{(m)}+x_{(m+1)}), & n=2m,\\
med= & x_{(m+1)},& n=2m+1\\

  \end{array} \right.
\end{displaymath}
The breakdown point for this estimate is 0.5, i.e.,  almost half
the data may be contaminated by outliers. But the effectiveness of
$T_{3}$ is much lower than for the $T_{1}$ and $T_{2}$: the
$LOSS=0.6$.

\subsection{Interquartile range}
This estimate is defined by
\begin{equation}
T_{4}=[(x_{1-q}-x_{q})/1.35]^2,
\end{equation}
where $q=0.25$ and for  distribution  $\cal{P}$  $x_{q}$ satisfies
$\cal{P}$$(x \leq x_{q})=q$. The breakdown point for $T_{4}$ is 0.25
and the $LOSS=0.6$. Fig. 1i shows the relative empirical influence
functions  for the variances computed with  $T_{4}$.

\subsection{Median  of pairwise averaged  squares}
This estimate is motivated by the Hodges-Lehmann estimate of
location
\begin{displaymath}
\widehat{\mu _{n}}=med(\frac{x_{i}+x_{j}}{2}, 1\leq i\leq j\leq n),
\end{displaymath}
which for variance is
\begin{equation}
T_{5}=1.483\times med_{1\leq i\leq jn}[(x_{i}^{2}+x_{j}^{2})/2]
\end{equation}

Fig. 1d shows the relative empirical influence functions  for the
variances computed with $T_{5}$  . The parameter $\epsilon$
(``eps'') indicates the percentage of outliers in the total volume
of data. The $LOSS$ for $T_{5}$ is 0.78.

\subsection{ $Qn$  estimate}
This estimate is proposed in \citep{rous93} and requires fewer
operations than $T_{5}$, but is quite effective. It combines the
ideas of the  Hodges-Lehmann estimate and the Gini estimate
\citep{kendal67}:
\begin{equation}
T_{6}=2.2219\{\left| x_{i}-x_{j}\right| ,i\leq j\}_{(k)},
\end{equation}
where  $k={h \choose 2}$ and $h=[n/2]+1, k \approx {n \choose 2}/4$.
This  estimate  is the $kth$ order statistic of the ${n \choose 2}$
interpoint distances. Fig. 1e shows the relative empirical influence
functions  for  variances computed with $T_{6}$. The $LOSS=0.88$.

\subsection{Biweight variance}
First the following auxiliary values are calculated
\begin{displaymath}
Y_{i}=\frac{x_{i}-M}{9\times MAD},
\end{displaymath}
where $M$ and $MAD$ are the median mean and median absolute
deviation, respectively. Then the coefficients are defined as
\begin{displaymath}
a_{i}=\left\{\begin{array}{ll}
 1, & if \left| Y_{i}\right| <1\\
 0, & if \left| Y_{i}\right| \geq 1
\end{array}\right.
\end{displaymath}
in which case the estimate is
\begin{equation}
T_{7}=\frac{\sqrt{\sum_{i=1}^{n}a_{i}(x_{i}-M)^{2}(1-Y_{i})^{4}}}{\left|
\sum_{i=1}^{n}a_{i}(1-Y_{i}^{2})(1-5Y_{i}^{2})\right| }
\end{equation}
Fig. 1f shows the relative empirical influence functions   for
variances computed with $T_{7}$. The $LOSS=0.92$.

\subsection{Bend Midvariance}
This estimate of variance is described in \citep{wilcox04}. Set
$\beta =0.1$ and  $m=[(1-\beta )n+0.5]$. Let $W_{i}= \left|x_{i}-M
\right|,i=1,...n$, and $W_{(i)}$ are these numbers sorted in
ascending order, M is the median mean of data $x$.
 Then the estimate of  $1-\beta$ quantile of the distribution of $W_{(i)}$ is $\widehat{\omega _{\beta}}=W_{(m)}$. Now set
\begin{displaymath}
Y_{i}=\frac{x_{i}-M}{\widehat{\omega _{\beta }}}
\end{displaymath}
and
\begin{displaymath}
a_{i}=\left\{\begin{array}{ll}
 1, & if \left| Y_{i}\right| <1\\
 0, & if \left| Y_{i}\right| \geq 1
\end{array}\right.
\end{displaymath}
in which case the bend midvariance is
\begin{equation}
T_{8}=\frac{\omega _{\beta }^{2}\sum_{i=1}^{n}\{(\psi
(Y_{i})\}^{2}}{(\sum_{i=1}^{n}a_{i})^{2}}
\end{equation}
where
\begin{displaymath}
\psi (t)=\max [-1,\min (1,t)].
\end{displaymath}
Fig. 1g shows the relative empirical influence functions  for
variances computed with $T_{8}$. The $LOSS=0.96$.

\subsection{Estimate with exponential weighting}
This estimate is described in \citep{shur01}. Mean is estimated as a
solution of the  equation
\begin{equation}
\widehat{\mu _{r}}: \hspace{0.5cm} \sum_{i=1}^{n}(x_{i}-\widehat{\mu
_{r}})e^{-q_{i}/4}=0
\end{equation}
and variance is defined as a solution of the equation
\begin{equation}
T_{9}=\sigma
_{r}^{2}:\hspace{0.5cm}\sum_{i=1}^{n}[(x_{i}-\widehat{\mu
_{r}})^{2}/\sigma _{r}^{2}-2/3]e^{-q_{i}/4}=0,
\end{equation}
where $q_{i}=(x_{i}-\widehat{\mu_{r}})^{2}/\widehat{\sigma ^{2}}$.
Fig. 1h shows the relative empirical influence functions     for
variances computed with $T_{9}$. The $LOSS=0.98$.

D. A. Lax performed a Monte Carlo study of more than 150 variance
estimators. Seventeen of these estimators were selected as being
either promising or commonly used \citep {lax85}. The results of our
computer simulation are in agreement with this test \citep {lax85},
especially concerning the high efficiency of  estimate $T_{1}$ and
$T_{7}$ and the low efficiency of  $T_{3}$.

The choice of a particular estimate depends on the type and
intensity of RFI, the type of observations and the method of
implementation (hardware or software).  $T_{8}$ and $T_{9}$  are the
best estimates from the point of view of  $LOSS$.  $T_{3}$ and
$T_{5}$ remove outliers in a most effective way (high  value of  the
breakdown point). Number sorting   and permutations of pairwise
measurements which are necessary in several algorithms require  more
computational time and computer memory.

\section{Computer simulations}

This section presents computer simulations of RFI mitigation using
algorithms which were employed in observations (Section 4).  Two
estimators from the the previous section have been chosen:
winsorization and exponential weighting. The reasons for this choice
are as follows.

All algorithms described in Section 2 give estimates of variance,
i.e., they work as  total power detectors (TPD) in radiotechnical
terminology. They can be applied in  single dish observations both
in continuum and in spectral observations. In the same manner  they
can substitute for TPDs   which are already  installed in  existing
radio telescopes.  Nowadays it is practically impossible for
technical and organizational reasons.In future  radio relescopes may
be equipped  with some RFI mitigation techniques and these variance
estimators could then be implemented in hardware or software shape.
But at the present time any experiment with RFI mitigation at
existing ratio telescopes must take  the
 technical constraints of implementation into account. Therefore only those estimators from Section 2 were chosen which could provide  not only  estimates of variance but also  ``clean'' data which could be applied further to TPD or to the correlator already in use in a radio telescope backend.

\subsection {Winsorization followed by total power detectors}
  Winsorization  with the parameter $\gamma=0.05$  was chosen
mainly because  impulse-like RFI in the temporal domain was
predominant during observations made at Effelsberg  presented in
Fig. 8. It was strong and sparse.  The percentage of  RFI in the
whole volume of data was not large $(<5\%)$.  Both  winsorization
and trimming can provide raw data without outliers but trimming
inconveniently reduces the number of samples   because of this
implementation problem: ``timing'' is distorted, i. e., there are
gaps (not zeros) in the presumed unbroken flow of data and the radio
telescope backend has no information about the location of these
gaps. The problem is easily solved in software during simulations
but may be critical in real observations. Winsorization  unlike
trimming preserves the total number of samples.

The block diagram  of computer simulation presented in Fig. 2  shows
the structure  of the algorithm using winsorization. The input
signal is the mixture of three signals: system  noise with normal
distribution ($\sigma=0.5$),  signal noise, also with normal
distribution ($\sigma=0.05$)  and impulse-like interference
imitating RFI. The signal noise is switched ``on'' and ``off''
emulating ``on-source'' or ``off-source'' position of the antenna
main beam, respectively. Poisson distribution ($\lambda=0.004$)
governs the appearance of RFI  and the lognormal distribution
(mean=12, standard deviation=6) determines the random amplitudes of
impulses. Input signal samples are stored in the buffer and sorted
to create  order statistics. The length of these statistics is equal
to $N=100$. The winsorization block provides both  variance
estimates for each of the $N$ samples and also the winsorized
samples which are applied to the external TPD (as in observations
presented  in subsection 4. 1). The loop with $M=100$ cycles
corresponds to  post-detector averaging: the  estimates of variance
after the winsorization block  are accumulated during $M$  cycles
and the mean calculated for these $M$  cycles represents the
simulation result.

Fig. 3    shows the results of computer simulations :\\
a) input noise with  normal distribution, $\mu=0.0, \sigma=0.5$, no
interference;
b) total power detector output, each point in this figure corresponds to squaring and averaging of  $N \times M=10^4$ samples in figure a); there are two steps, ``up'' at point $\#100$ and  ``down'' at point $\#200$  corresponding to the increase of $\sigma$  from value $0.5$ to the value $\sigma+\Delta \sigma, \Delta \sigma=0.05$ ( ``on-source'' and ``off-source'' position of antenna beam);\\
c) interferences in the form of random impulses are added to the noise a);\\
d) total power detector output with the input signal  c);\\
e) total power detector output with the input signal c) and
preliminary winsorization.  The difference between the scales in the
vertical axes  in d) and e)  demonstrates the effect of interference
mitigation.

 \subsection{Exponential weighting}
Exponential weighting  also provides not only  estimates of variance
but  data with suppressed outliers. This estimator is also of
superior  effectiveness. And unlike winsorization, it does not sort
input data. The influence of sorting on  phase information can be
detrimental to radiointerferometic  and pulsar (de-dispersion
procedure) observations. Impulse-like RFI in the frequency domain
was prevalent during the pulsar and image synthesis observations at
WSRT (subsection  4. 2 and 4. 3).  It is to  demonstrate
exponential weighting in the frequency domain which is the aim of
this computer simulation.

The block diagram  presented in Fig. 4 shows the structure of the
algorithm using exponential weighting in the spectral domain. The
input signal is a mixture of four signals: system  noise with normal
distribution ($\sigma=1.0$) filtered to emulate the WSRT backend
low-pass filters; two noise-like signals with normal distribution
imitating emission and absorption spectral lines, respectively, and
interference waveforms. A randomly phase-modulated  sinusoidal
carrier  is used as the RFI signal, the index of modulation is equal
to $\pi /2$ (binary phase modulation). Poisson distribution
($\lambda=0.3$) determines  the moments of phase jumps.

There are two loops in the algorithm: the inner loop ($m=1,..M$)
which is responisble for the storage of $M$ instantaneous spectra of
the input signal for RFI mitigation processing, and the outer loop
($l=1,..L$) which repeats  consecutively $L$ identical stages of
processing inside the inner loop,  showing the  dynamics of the
running spectrum.  Interference is intermittently switched  ``on''
and ``off'' inside the inner loop, the percentage in the total
averaging interval is less than $7\%$.

$N$ samples of the input signal are Fourier-transformed with
forward FFT, $N=512$. The inner loop with the counter ($m=1,..M$)
provides $M$ complex spectra (each having a length equal to $N$)
which are stored in the buffer.  So there are two sets (real and
imaginary) of data in the buffer numbered  as $m=1,..M$ for each of
the $N=512$ spectral channels. $L$ power spectra of the input signal
averaged on $M$ instantaneous spectra are calculated  in the outer
loop ($l=1,..L$). Fig. 5a shows the three-dimensional time-frequency
presentation of  the sequence of $L=50$  averaged ``dirty'' spectra.

 Equation (18) is solved separately for each of $2 \times N$  real and imaginary sets of data providing estimates of $\widehat{\sigma^2_{M,real}(i)}$ and  $\widehat{\sigma^2_{M,imag}(i)}$  - variances in $i=1,..N$ spectral channels calculated using $M$ samples of  real and imaginary components of the instantaneous spectra. Random values in the real and imaginary part of the instantaneous spectrum are independent, so the sum of these  estimates in each spectral channel gives the total estimated power spectrum for each spectral channel.  This ``clean'' power spectrum is represented in Fig. 5b for $L=50$ time intervals.
In real observations each of  $l-th$ time interval is equal to $N \times M \times \Delta t$ seconds, where $\Delta t=1/2\Delta f$ is the input signal sample interval,  $\Delta f$ is the bandwidth of the input signal.  \\
 In  applications  it is practical  to find  the solution to  (18)  when $\mu_{r}=0$ using the  approach of stochastic approximation:
\begin{equation}
\widehat{\sigma _{m}}=\widehat{\sigma
_{m-1}}+\frac{1}{m}\sum_{k=1}^{m}(\frac{x_{k}^{2}}{\widehat{\sigma
_{m-1}}}-\frac{3}{5})\exp (-\frac{x_{k}^{2}}{3\widehat{\sigma
_{m-1}}}),
\end{equation}

$m=1,..M$. Fig. 5c and 5d show the result of averaging   $L$
spectra  in Fig. 5a and 5b for the ``dirty'' and ``clean''power
spectra, respectively. The frequencies of interferences do not
coincide with the frequencies of spectral lines . It can therefore
be seen that the components with  normal distribution (system noise
and spectral lines)  are untouched by this RFI mitigation procedure.
The sequence of pictures in Fig. 6 similar to Fig. 5 demonstrate the
result of computer simulation when the frequencies of interferences
coincide with the frequencies of spectral lines .  In this case  we
again see   that the restoration of the spectra is  satisfactory.

Figures 5 and 6  demonstrate  how the algorithm works with  spectra.
However  experimental constraints did not require the provision of
estimates of variance or power spectrum but did required the
``clean'' signal in  the temporal domain similar to the input signal
(the level of the signal, the bandwidth). Therefore  running
estimates of   ``clean'' and ``dirty'' power spectra are used for
exponential weighting of the running complex instantaneous spectra:
each complex value of an instantaneous spectrum being multiplied by
$exp(-P(i)/3\widehat{\sigma^2(i)})$, where $P(i), i=1,..N$ is the
input spectral variance (power spectrum) in the channel  $i$,
$\widehat{\sigma^2(i)}$ is the estimate of the ``quiescent'' power
spectrum in the channel  $i$ found from Eq. (18). A corresponding
delay must be introduced because of the time required  to calculate
all $\widehat{\sigma^2(i)}$. Then, after the backward FFT, the
signals in the temporal domain can be applied to TPDs or
correlators.  This auxiliary output was used to supply the ``clean''
raw data in the observations described in Sections 4. 2  and 4. 3.

Figure 7 shows the impact of RFI mitigation on cross-correlation
(radiointerferometric observations). The algorithm in the block
diagram  shown in Fig. 4 was applied to two signals with  the
additional coherent Gaussian component ($\sigma_{s}=0.1$) emulating
the noise from a radio source  received at both sites of  the radio
interferometer. Sporadic interference with the duty cycle equal to
$0.2$ was generated as  frequency-modulated carrier and was
identical at both sites which is the worst case scenario: $100\%$
correlated RFI. In reality the impact of  RFI in radio
interferometers is considerably reduced  due to fringe stopping and
delay tracking procedures \citep{thom82}.

 Figures 7a and 7b demonstrate  three-dimensional time-frequency presentations of the input ``dirty''
and ``clean'' power spectra on one site, respectively. Deep
``troughs''   can appear in the spectrum in Fig. 7b  at the places
of interferences due to the exponential effect of weighting: when
interference is strong the algorithm works similarly to the
``thresholding and blanking'' algorithm but more smoothly and
without  the {\it a priori} knowledge  necessary for the positioning
of the threshold level. Figures 7c and 7d give the averaged
normalized cross-correlation functions corresponding to Fig. 7a and
7b, without and with RFI mitigation, respectively. The central parts
of the cross-correlation functions between channels $\# 400$ and
$\#600$ are shown. The scale of the vertical axe in Fig. 7c is
significantly larger   than that in Fig. 7d thus showing strong
excessive cross-correlation due to RFI.

It is necessary to add two further comments here for the purposes of
discussion.

1. The  estimators of  variance described in Section 2 and those
tested in Section 3  are essentially nonlinear procedures.  For
example, one can, at least theoretically, imagine a  huge  outlier
which  renders  the conventional estimator (1) completely
non-functional whereas these estimators work perfectly.
Radioastronomers  often ask what is the value of RFI suppression
(usually  in dB) when an RFI mitigation algorithm is applied?
Example in Fig. 3 demonstrates that it is not always possible to
answer this question correctly. The larger  the outlier's amplitudes
the more effectively they are deleted. If the amplitudes of impulses
in Fig. 3c are 10 times larger the result in  Fig. 3e will be the
same or even slightly improved. This is also valid for the
``thresholding and blanking'' algorithm which has been succesfully
used in several articles on RFI mitigation. That is why  it is
better to judge an RFI mitigation procedure in each paticular case
in a combined manner, that is looking at the ``signal-of-interest''
distortions by both the RFI residuals and the procedure itself.

2. The estimators of Section 2 are intended for  applications when
sporadic, impulse-like interference disturbs observations. However
interference
 is often continuous, persistent and practically constant during the averaging interval.
 For example, for $M=100, N=512, \Delta f=20MHz$ this interval is equal to $1.28 \times 10^{-3}$ sec which is rather a short  interval
for several RFI to change their amplitudes. The estimators of
Section 2 will not yield any benefits in this case. But the same
fact that RFI is quasi-constant at a reasonably  chosen averaging
interval  can help    to decouple the Gaussian component with a
``quiescent'' value of $\sigma$ and RFI with ease. A simple
procedure can do this. The power spectrum after FFT is usually
calculated as

 \begin{equation}
P(f)=1/M\sum_{m=1}^{M}\{Re[s(f)_m]^{2}+Im[s(f)_m]^{2}\}
 \end{equation}
The sums of RFI and ``useful'' Gaussian noise in each spectral
channel are squared and averaged. Strong RFI dominates in this case
over the noise. Another algorithm is proposed here: to  separately
estimate the variances  of real and imaginary parts of the spectrum
and then add them to obtain  the total variance in each spectral
channel, i.e., the power spectrum sought. The variances of real and
imaginary parts of the complex spectrum must be calculated taking
into consideration that the mean value now is no longer  equal to
zero because of the presence of RFI.  So the general formula for
sample  variance must be applied: for a random value $x$, the sample
variance is $var(x)=$$\widehat{\sigma
_{x}^{2}}=1/M\sum_{m=1}^{M}(x_{m}-\widehat{x})^{2}$ which for the
power spectrum at frequency $f$ is

\begin{eqnarray}
P(f)=var[Re(s(f)]+var[Im(s(f)]=\\
1/M\sum_{m=1}^{M}Re[s(f)]_{m}-\widehat{Re[s(f)}]\}^{2}+\nonumber\\
1/M\sum_{m=1}^{M}Im[s(f)]_{m}-\widehat{Im[s(f)}]\}^{2},\nonumber\\
\widehat{Re[s(f)}]=1/M\sum_{m=1}^{M}Re[s(f)]_{m},\nonumber\\
\widehat{Im[s(f)}]=1/M\sum_{m=1}^{M}Im[s(f)]_{m}\nonumber
\end{eqnarray}

In this case  RFI is eliminated due to its  constant value  for all
$M$ samples of the instantaneous spectra $s(f)]_{m}$.  It may also
be useful for the mitigation of  weak but persistent RFI whose
detrimental impact is revealed only after lengthy averaging. This is
a conjecture which has been proved in computer simulations but
which, of course, requires
 experimental confirmation.

\section{Examples from observations}

\subsection{ Observations in continuum at Effelsberg  radio telescope}

Radio source 1448+762 was observed in continuum at the Effelsberg
radio telescope. The receiver  output  was split on two channels.
The signals from one channel were processed  by  a FPGA processor
(Altera Stratix S80) and then  sent  to the total power detector.
The signals from the second channel  were applied straight to the
total power detector. The bandwidth of the signals applied to the
total power detector was equal to 20 MHz. These  channels provided
two radio telescope outputs: one  output with RFI mitigation and
another  without. Pairs of radio source scans from these outputs
were made simultaneously.  Because the  total power detector  was an
integral part of the radio telescope backend equipment,  the
algorithm implemented in FPGA  only processed the IF (intermediate
frequency) signal with the aim of ``cleaning" it of  RFI. The
analogue input signal was digitized in 12-bit ADC with  40Msamples
speed, then processed in FPGA,  transformed back into  analogue form
and applied to the total power detector. Winsorization of the signal
in temporal domain ($\gamma=0.05$) was implemented.

Fig. 8 shows  eight scans of the  radio source, each scan
represented by two panels: the top panel - scan with RFI mitigation,
the lower panel - without RFI mitigation. Fig. 9  displays radio
images  of the source  built using  scans similar to those in Fig.
8:  the left panel - without RFI mitigation, the right panel - with
RFI mitigation.

\subsection{Pulsar observations  at WSRT}

A new pulsar machine PUMA-2 has been installed at the Westerbork
synthesis radio telescope (WSRT). The radio telescope works in
tied-array mode in which all 14 signals from antennas are added in
phase , i.e., there is one   output  (in reality with  two
polarizations) as for a single dish. The 20 MHz-baseband signals
from each of the eight  frequency channels of WSRT  are digitized (8
bit) and stored in the mass storage system which has sufficient
hard disk capacity to support at least 24 hours of continuous
observations. Signal processing can  therefore be undertaken  {\it
off-line}.

In our experiment a block of  data  recorded during 10 sec, $40
\times 10^6$  eight-bit samples/sec, was used. All RFI mitigation
processing and the total power detector (TPD) were realized entirely
in software during {\it off-line} processing.  The estimate of
variance with exponential weighting was used. Fig. 10  displays the
results of processing. Upper row, left panel: TPD outputs for  two
polarizations calculated from the  raw data with RFI, right panel:
TPD outputs, RFI removed. Middle row, left panel: example of a time
fragment  of the power spectrum with RFI;  right panel : the same
time fragment, RFI removed. Lower row shows pulsar profiles after
de-dispersion and  folding at the pulsar period, left panel: pulsar
profile  averaged  over 10 sec from both polarizations of raw data,
middle panel: similar pulsar profile, RFI removed, right panel:
pulsar profile restored with observational data obtained at 1420 MHz
without any RFI which is put here for comparison with the profile in
the middle panel.

\subsection{Radio image synthesis at WSRT}

Radio source DA240 was observed at WSRT at  a central  frequency of
357 MHz with the  bandwidth  equal to 20 MHz in  the presence of
strong RFI. The RFI mitigation system (RFIMS) was used for {\it
real-time} processing \citep{baan04}. Analogue baseband signals
were digitized (12bit ADC, 40 Msamples/sec), processed in FPGA
(Altera StratixS80) and  transformed back to analogue form for
subsequent processing in the WSRT correlator. The algorithm used for
the removal of RFI  was similar to the algorithm with exponential
weighting, except that the variance was not used as an output,
instead   ``cleaned" , exponentially weighted  signals were applied
to the correlator.

Radio images of the source DA240 are shown in Fig. 11. Upper row,
left panel: image without RFI mitigation; right panel: image with
RFI mitigation. Lower row:  central parts of  the image presented in
the same order.  The stretched form of the synthesized images is
explained by the fact that the observations lasted  8 hours, instead
of a full 12 hour  aperture synthesis cycle.

RFI mitigation was implemented on each of 14 radio telescopes at
WSRT. This may give rise to some distortions because of the
difference of equipment  characteristics at the different antennas.
Other observations were made specifically to judge the  ``toxicity"
of the RFI mitigation procedure. Two pairs of radio images of  radio
source 4C34.47 were synthesized: one was observed at 1420 MHz with
the RFI mitigation system and without it and another at 345 MHz,
also with the RFI mitigation system and without it. The first pair
of images served as a test for ``toxicity" while the second pair
demonstrated the effectiveness of RFI mitigation. Both pairwise
observational data  were stored during  simultaneous observations.

Fig. 12 displays the results of the image synthesis.
 Upper row, left panel: central  frequency 345 MHz, without the RFI mitigation system; right panel: central  frequency 345 MHz, with the RFI mitigation system. Notice the difference of  intensity levels in the figures.
Lower row, left panel: central  frequency 1420 MHz without the RFI
mitigation system.
 right panel: central  frequency 1420 MHz with the RFI mitigation system. After subtracting one image from the other the {\it rms} noise  is  less than 0.7 mJy/beam which signifies a good similarity  in the images.

\section{ Conclusions}

1. Statistical analysis of raw data with the finest available time
and frequency resolution can help during observations in an RFI
contaminated envinronment. Estimates of variance are an  important
part of both  classical  and robust  statistics. Growing concern
about RFI pollution  should persuade  the  radio astronomy community
to pay more attention to a variety of algorithms developed in the
realm of robust  statistics. Tradional radiometers with simple
square-law detectors, spectrum analyzers and correlators must be
equipped with these tools. This framework  of robust estimates puts
the successfully tested  blanking of RFI on a more stable
foundation.

2. Statistically faithful, robust  estimates of variance are
especially appropriate for application in an  impulse-like strong
RFI  environment. RFI is effectively  suppressed and the
accompanying  loss in the signal-to-noise ratio is  tolerable. Table
3 gives  a  summary  of available RFI mitigation methods which have
been proposed during the last years. Some of these have been tested
in real observations.   Blanking showed truly good results in the
case of    impulse-like strong RFI. There are slots in the Table 3
where  ``removing or blanking""  is referred to. The aforementioned
robust  algorithms can be  applied in these particular situations.

3. The results produced from the  observations at the Effelsberg
radio telescope (image synthesis by beam scanning) and at WSRT
(Earth rotation aperture synthesis) included  in this article
demonstrate the usefulness of this kind of RFI mitigation both in
{\it real-time} and  {\it off-line}  (pulsars).

4. The choice of a particular algorithm depends on the type and
intensity of RFI.The proportion of RFI  presence in data is also
important. The type of implementation may determine the choice:
{\it off-line}  or {\it real-time}.  All options are open now. Mass
storage systems allow data  to be processed {\it off-line} on
powerful computers. Nowadays both existing radio telesopes  future
projects (LOFAR, ATA, SKA)  will generate such huge amounts of data
that  {\it real-time} processing is  vital: DSP, FPGA or
supercomputers are possible solutions. Therefore  the creation of
`` robust" radiometers, spectrum analyzers and correlators  is an
urgent necessity.




\acknowledgments I am  grateful to Juergen  Neidhoefer  and  Ernst
Fuerst for their help during observations and data processing at the
Effelsberg radio telescope, to Ben Stappers and  Ramesh Karuppusamy
for providing the pulsar data and  Subhashis Roy  for AIPS
processing of 4C34.47 data.

\clearpage \onecolumn
 \centerline{Table 3: The application of RFI mitigation methods}

\hspace{1cm}

\begin{tabular}{|l |l |l||l|l|}
\tableline
  Type of radio  & Type of RFI, & Type of RFI,   & Observations  in continuum & Spectral observations \\
  telescope & intensity  & structure  &   &   \\
\tableline \tableline

  &  & impulse-like &   tolerable & tolerable\\
\cline{3-5}
  & weak &  narrow-band & Removing (blanking)  & ANC with a reference channel \\
  & RFI &  &              in spectral domain & and post-TPD subtraction; \\
& & & & HOS analysis\\
\cline{3-5}

  & & wideband & ANC with a reference channel   &ANC with a reference channel \\
Single &  &  &        and post-TPD estimation & and post-TPD estimation \\
 & & &  \& subtraction  &  \& subtraction\\
 \cline{2-5}
dish &   & impulse-like & Removing (blanking) & Removing (blanking)\\
 & & &                   in temporal domain & in temporal domain\\
\cline{3-5}
  & strong  & narrow-band & Removing (blanking)& ANC with a reference\\
 & RFI& &                   in spectral domain &  channel; HOS analysis \\
\cline{3-5}
 & & wideband &  ANC with &  ANC with\\
& & &           a reference channel & a reference channel\\
\tableline
 &  &  impulse-like & tolerable & tolerable\\
\cline{3-5}
 & weak  & narrow-band &  tolerable & Post-correlation ANC with\\
  & RFI   &   & & reference channels; HOS analysis\\
\cline{3-5}
 & &wideband & tolerable & Post correlation ANC with\\
Connected   & & & & reference channels\\
\cline{2-5}
radio &strong & impulse-like & Removing (blanking)  & Removing (blanking) \\
interferometer  & RFI &  & in temporal domain & in temporal domain\\
\cline{3-5}
 & &  narrow-band & Removing (blanking) & ANC with a reference\\
& & & in spectral domain & channel. Spatial filtering\\
\cline{3-5}
& & wideband & ANC with a reference & ANC with a reference\\
& & & channel. Spatial filtering& channel.  Spatial filtering\\
\tableline
& weak & impulse-like & tolerable& tolerable\\
\cline{3-5}
 & RFI & narrow-band& tolerable& tolerable\\
\cline{3-5}
 &  & wideband &   tolerable& tolerable\\
\cline{2-5}
VLBI & strong & impulse-like & Removing (blanking) & Removing (blanking)\\
   & RFI  & &      in temporal domain &in temporal domain\\
\cline{3-5}
 & & narrow-band& Removing (blanking)& ANC with a reference\\
  & &  & in spectral domain & channel\\
\cline{3-5}
 & & wideband & ANC with a reference & ANC with a reference\\
& & &  channel & channel\\
\tableline
\end{tabular}

RFI -  radio frequency interference.\\
TPD - total power detector.\\
ANC - adaptive noise cancellation.\\
HOS - higher order statistics.\\

\clearpage

\onecolumn \centerline{\bf List of captions}

{\bf Fig. 1}. a) The empirical influence function as a function of RFI amplitude, $\sigma=1,n=10^{5}$, the estimates are made with initial data. The curves here and elsewhere are parameterized  by the fraction of RFI in the total volume of data:    $\epsilon=0.01;0.025;0.5$;\\
 b)  estimates  made using {\it trimmed} data; \\
 c)  estimates  made using {\it winsorized} data;\\
d)  estimates  made  as the {\it median of pairwise averaged  squares}; \\
 e)  estimates  made using the {\it $Qn$  estimate} algorithm; \\
 f)  estimates  made using the {\it biweight variance} algorithm; \\
g)  estimates  made using the {\it bend midvariance} algorithm; \\
 h)  estimates  made using the algorithm of {\it exponential weighting};\\
 i)  estimates  made using the algorithm of {\it interquartile range}.

{\bf Fig. 2}. Block diagram of computer simulations: winsorization
is used for RFI mitigation when  the total power detector is the
backend output as in section 4. 1.

{\bf Fig. 3. } Results of computer simulations with the algorithm shown in Fig. 2:\\
a) noise with the normal distribution, $\mu=0.0, \sigma=0.5$, no interference;\\
b) total power detector output, each point in this figure corresponds to squaring and averaging of  $10^4$ samples in figure a), there are two steps, ``up'' at point $\#100$ and  ``down'' at point $\#200$  corresponding to the increase of $\sigma$  from value 0.5 to the value $\sigma+\Delta \sigma, \Delta \sigma=0.05$.\\
c) interference is added to the noise a): random impulses with the
Poisson distibution ($\lambda=0.04$) and the lognormal
disrtibution of amplitudes (mean=10, standard deviation=5);\\
d) total power detector output with input signal  c);\\
e) total power detector output with  input signal c) and preliminary
winsorization (equation (9)),  note the difference of scale in d)
and e).

{\bf Fig. 4}. Block diagram of  computer simulations: exponential
weighting is used for RFI mitigation  in the frequency domain
when the total power detector is the backend output as in section 4. 2 (pulsar observations)\\
 or the correlator as in section 4. 3.
(radiointerferometric observations).

{\bf Fig. 5. } Results of computer simulations  of exponential weighting with the algorithm shown in Fig. 4:\\
a) time-frequency presentation of power spectrum consisting of
system noise, emission and absorption lines and RFI
(randomly binary-phase manipulated signals), L=50 time sections of spectrum divided on 256 channels, each spectrum is the mean of $M=100$  instantaneous spectra; \\
b) power spectrum obtained as a solution to equation (18) for each spectral channel and $M=100$  samples;\\
c) power spectrum  averaged in time, using $L=50$ sample spectra from a);\\
d) power spectrum  averaged in time, using $L=50$ sample spectra
from b), i.e., with RFI mitigation.

{\bf Fig. 6. } Results of computer simulations  of exponential weighting with the algorithm shown in Fig. 4:\\
spectra  of interference  overlap with  spectral lines. Other parameters are similar to Fig. 5,\\
a) time-frequency presentation of the power spectrum consisting of
system noise, emission and absorption lines and RFI;
b) power spectrum obtained as a solution to equation (18) for each spectral channel and $M=100$  samples;\\
c) power spectrum  averaged in time, using $L=50$ sample spectra from a);\\
d) power spectrum  averaged in time, using $L=50$ sample spectra
from b), i.e., with RFI mitigation. Both spectral lines are clearly
visible.

{\bf Fig. 7. } Results of computer simulations  of exponential
weighting with the algorithm shown in Fig. 4 when
``dirty'' and ``clean'' signals are applied to the correlator; L=50 time sections of the spectrum divided on 256 spectral channels, each spectrum is the mean of $M=100$  instantaneous spectra; the spectra at the first input of the correlator are shown, the   spectra at the second  input are similar to a) and b):\\
a) time-frequency presentation of power spectrum consisting of system noise, and RFI (frequency-modulated bursts);\\
b) power spectrum after RFI mitigation - exponential  weighting of
each instantaneous spectrum using variances
obtained as a solution to equation (18) for each spectral channel and $M=100$  samples;\\
c) cross-correlation in the presence of RFI and no RFI mitigation, the central 200 channels are shown;\\
d) cross-correlation in the presence of RFI and with RFI mitigation; \\
take notice of the difference of the vertical scales in c) and d).

{\bf Fig. 8}. Examples  of  RFI mitigation at the Effelsberg radio
telescope during observations in continuum at central frequency 1645
MHz, bandwidth 20MHz. A selection of  eight scans  of the source
1448+762 is represented. Pairwise records were made simultaneously
for the channel with RFI  mitigation  and the channel without RFI
mitigation.

{\bf Fig. 9}. Radio image of the source 1448+762 built using scans
similar to those in Fig. 2:  left panel - without RFI mitigation,
right panel - with RFI mitigation.

{\bf Fig. 10}.  Pulsar B0329+54.07 observed at WSRT at 1625MHz. Data
were recorded during 10 sec, 40 Msamples/sec. Upper row, left panel:
TPD ouputs for two polarizations, raw data with RFI, right panel:
TPD output, RFI removed. Middle row, left panel: time fragment  of
the running power spectrum with RFI; the same time fragment, RFI
removed. Lower row, Left panel: pulsar profile made with raw data
over 10 sec, middle panel: pulsar profile, RFI removed,  right
panel: pulsar profile observed at 1420 MHz, no RFI.

{\bf Fig. 11}. Radio images of the source DA240 observed at WSRT,
central  frequency 357 MHz, bandwidth 20 MHz.  Upper row, left
panel: image without RFI mitigation; right panel: image with RFI
mitigation. Lower row:  central parts of  the images  with and
without RFI.

{\bf Fig. 12 }. Radio images of the source 4C34.47 observed at WSRT,
bandwidth 20 MHz.
 Upper row, left panel: central  frequency 345 MHz, without RFI mitigation system; right panel: central  frequency 345 MHz, with RFI mitigation system. Notice  the difference of the intensity levels in the figures.
Low row (``toxicity" test), left panel: central  frequency 1420 MHz
without RFI  mitigation system.
 right panel: central  frequency 1420 MHz with RFI mitigation system. After subtracting one image from the other the {\it rms} noise  is  less than 0.7 mJy/beam which signifies a good similarity  in the images.



\begin{figure}
\includegraphics[width=5.5cm,height=5.5cm]{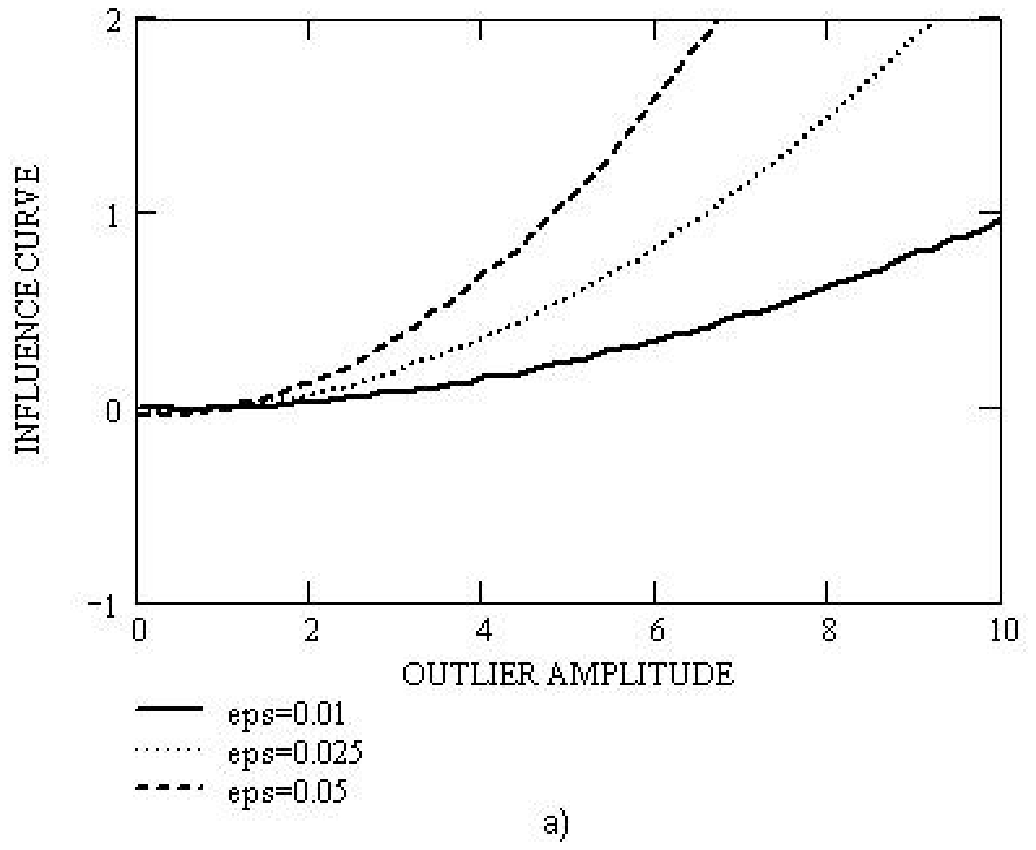}
\includegraphics[width=5.5cm,height=5.5cm]{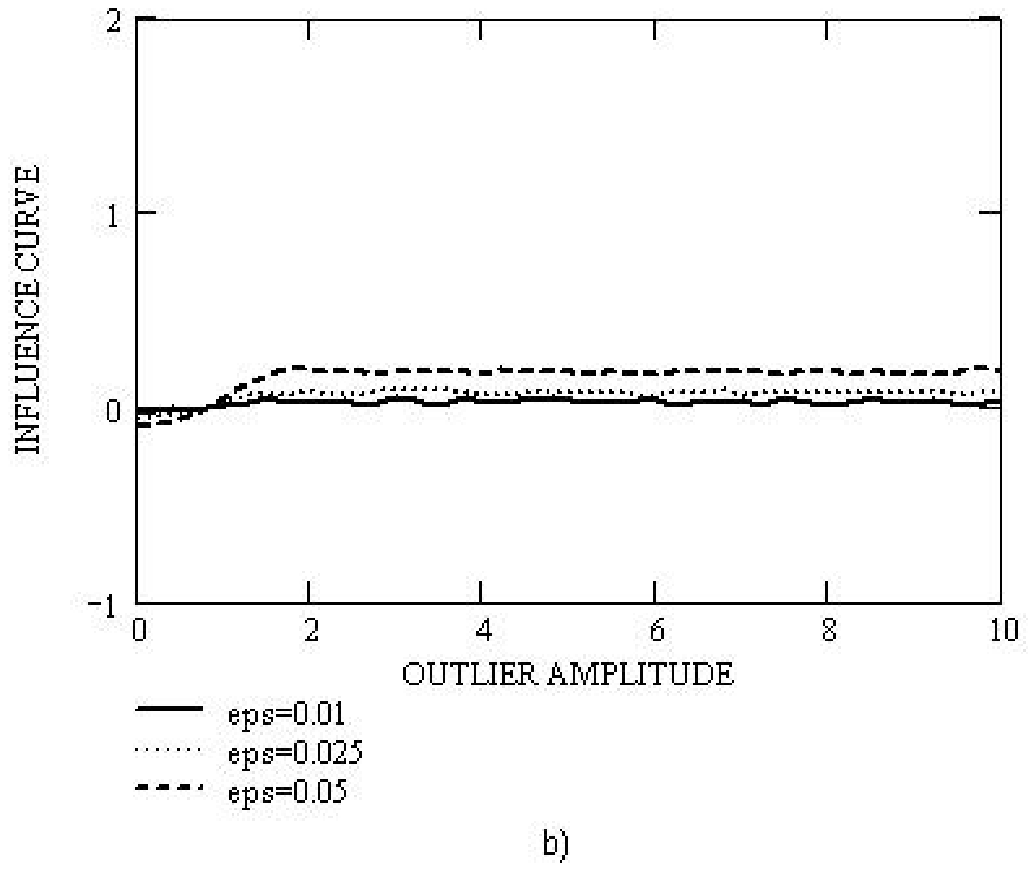}
\includegraphics[width=5.5cm,height=5.5cm]{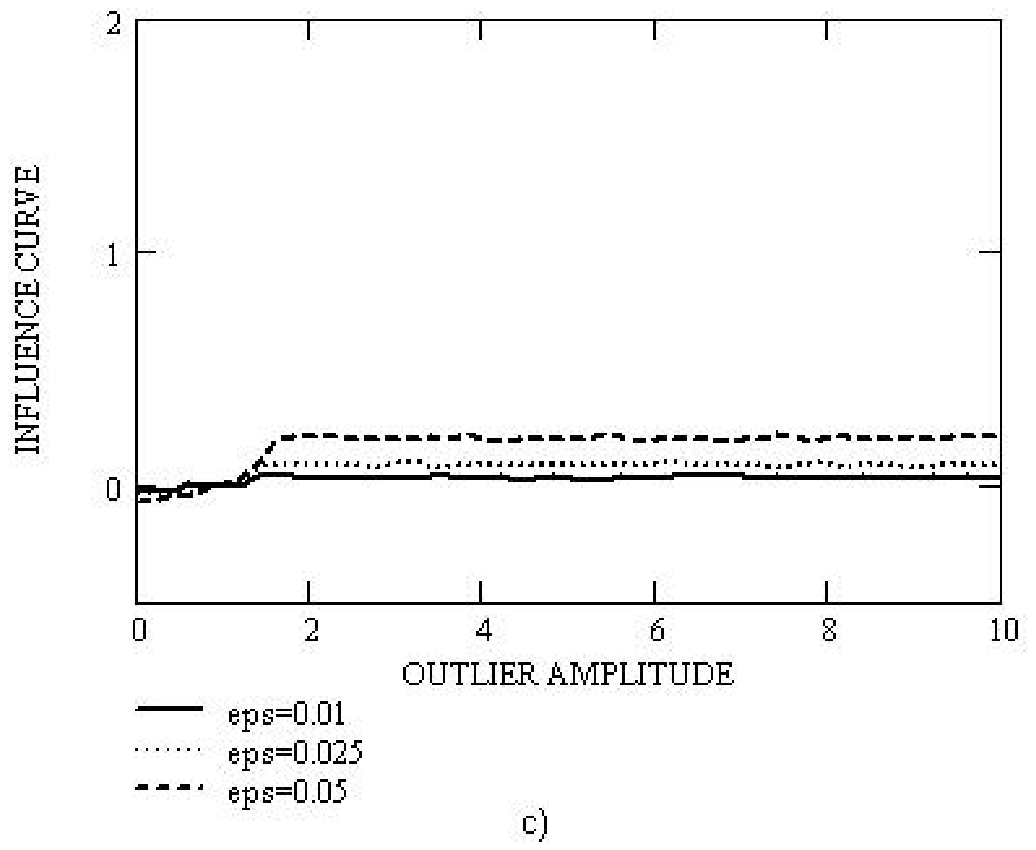}
\includegraphics[width=5.5cm,height=5.5cm]{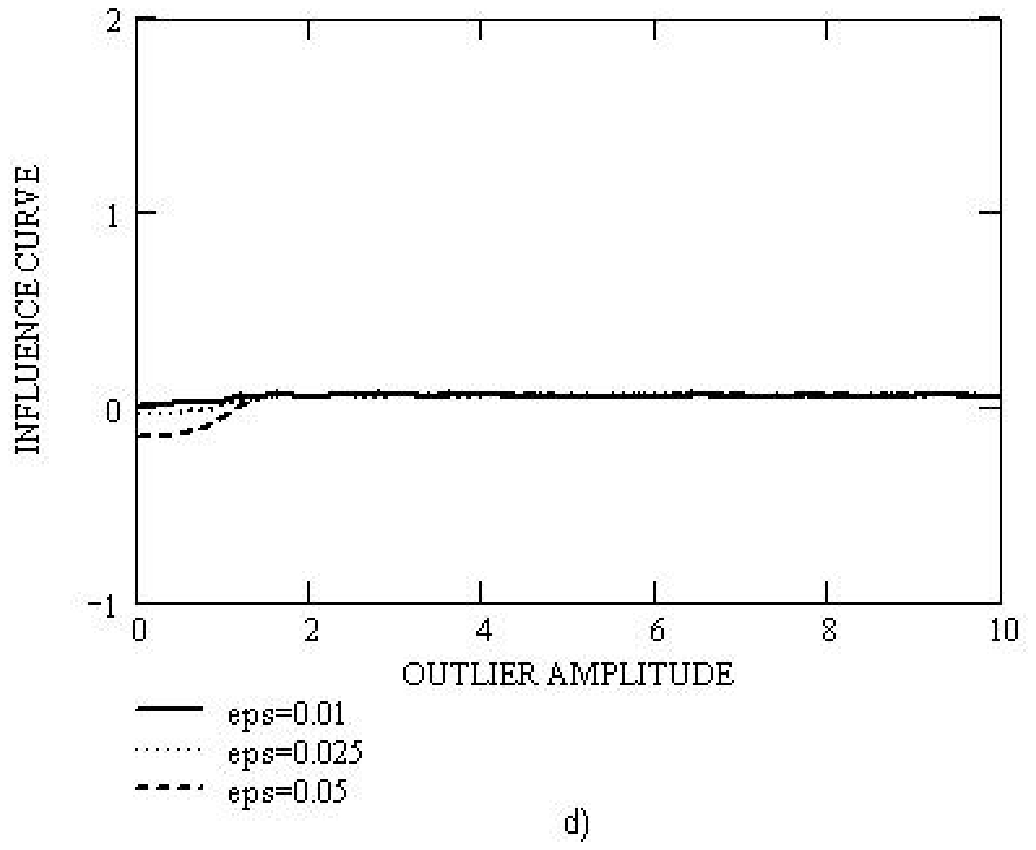}
\includegraphics[width=5.5cm,height=5.5cm]{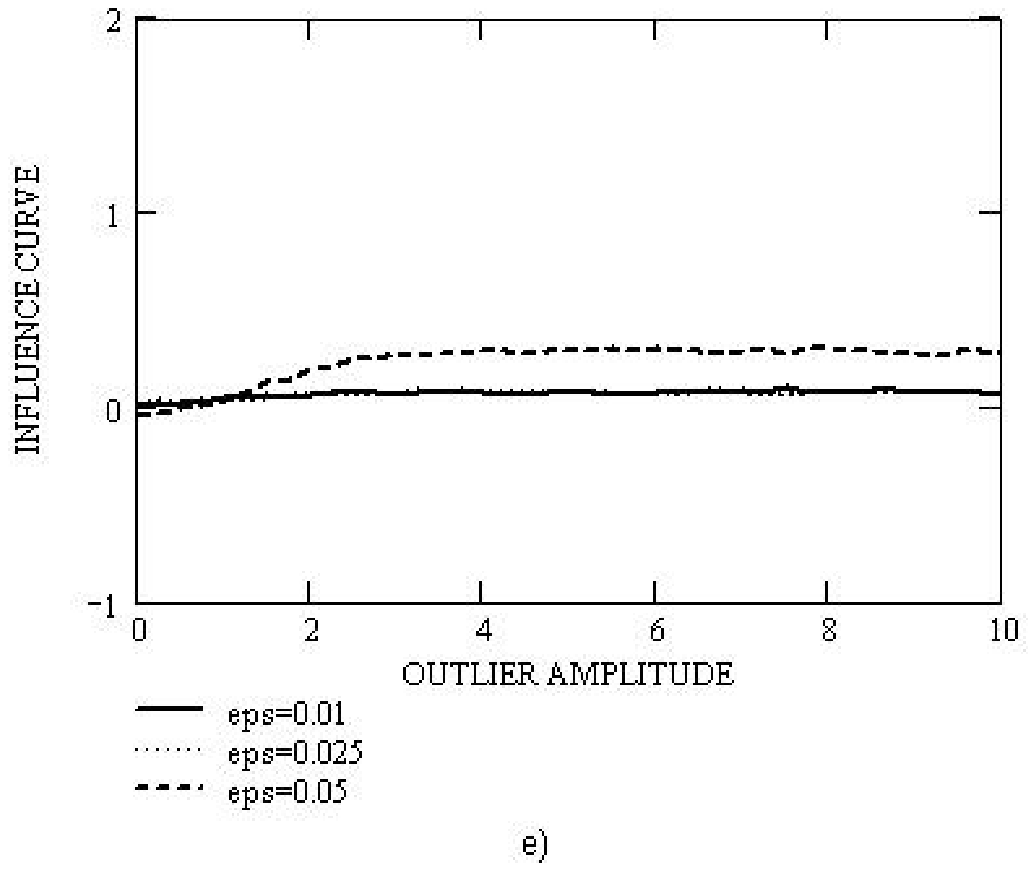}
\includegraphics[width=5.5cm,height=5.5cm]{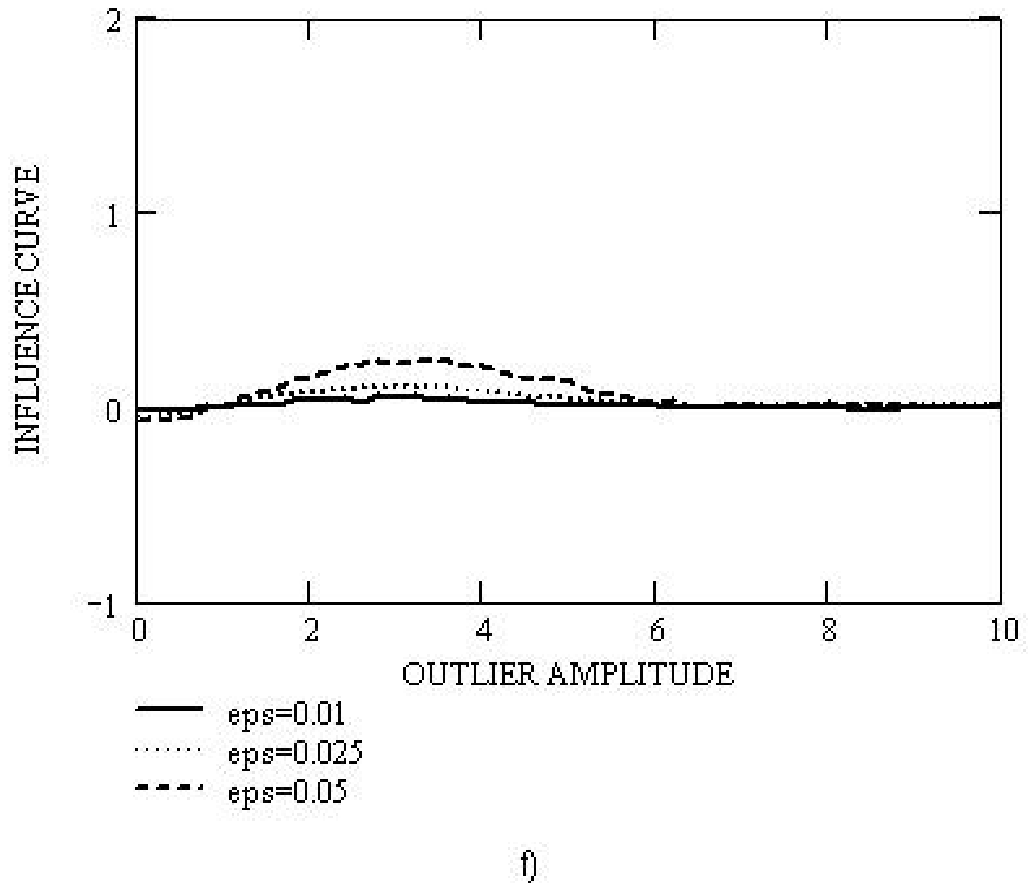}
\includegraphics[width=5.5cm,height=5.5cm]{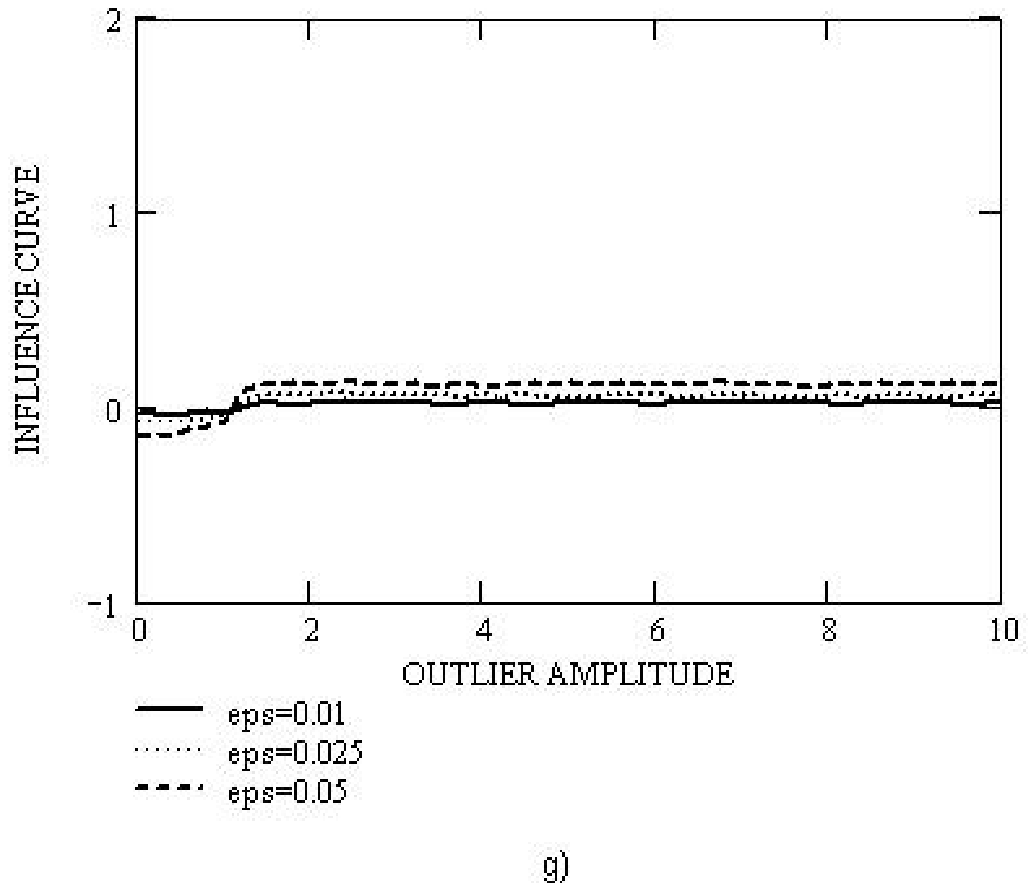}
\includegraphics[width=5.5cm,height=5.5cm]{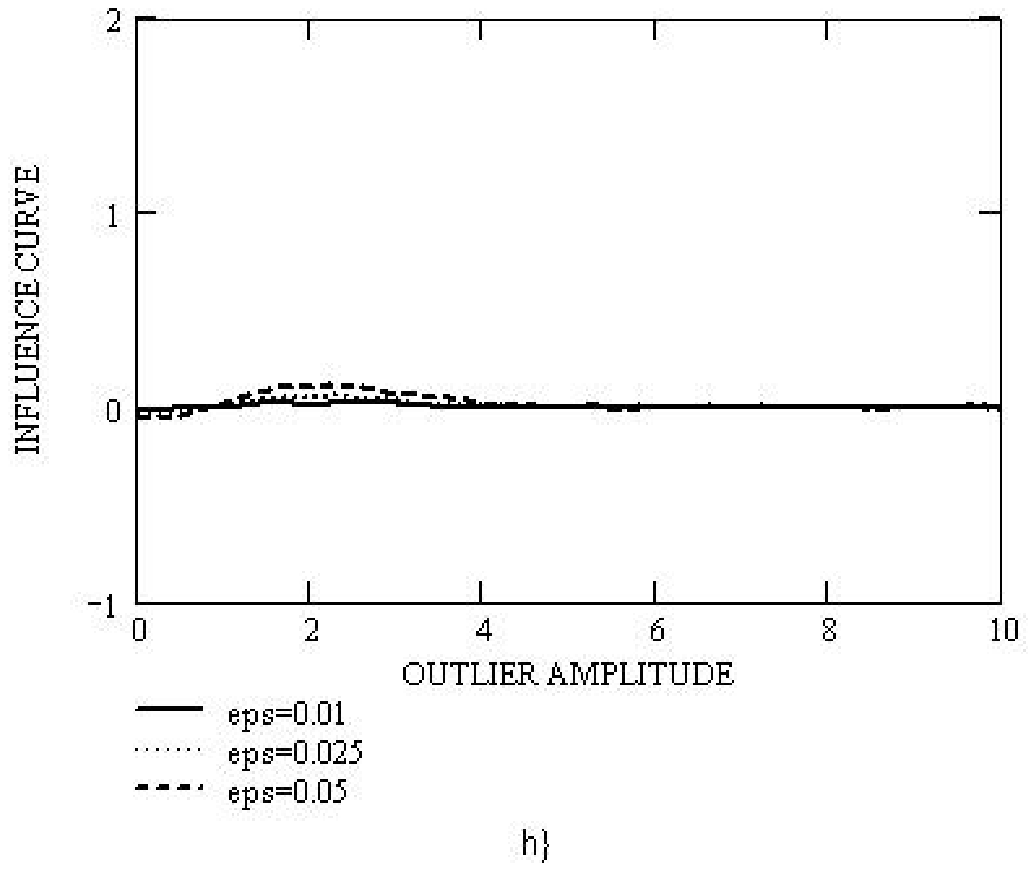}
\includegraphics[width=5.5cm,height=5.5cm]{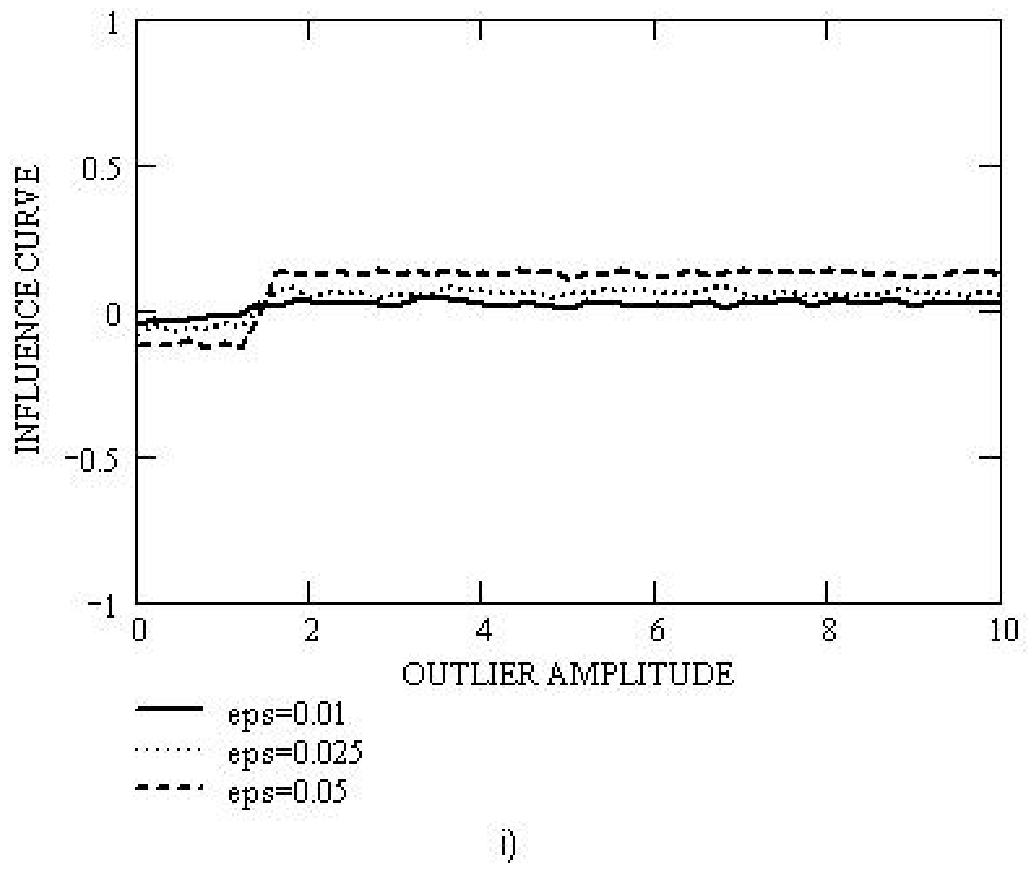}

\caption{a) The empirical influence function as a function of RFI
amplitude, $\sigma=1,n=10^{5}$, estimates  made with initial data.
The curves here and elsewhere are parameterized  by the fraction of
RFI in the total volume of data:$\epsilon=0.01;0.025;0.5;$
 b) estimates  made using {\it trimmed} data;
c)  estimates  made using {\it winsorized} data; d)  estimates  made
as the {\it median of  pairwise averaged  squares };
 e)  estimates  made using the {\it $Qn$  estimate} algorithm;
 f)  estimates  made using the {\it biweight variance} algorithm;
g) estimates  made using the {\it bend midvariance} algorithm;
 h)  estimates  made using the algorithm of {\it exponential weighting} . }
 i) estimates  made using the algorithm of {\it interquartile range}.
\end{figure}

\begin{figure}
\includegraphics[width=10.5cm,height=15.5cm]{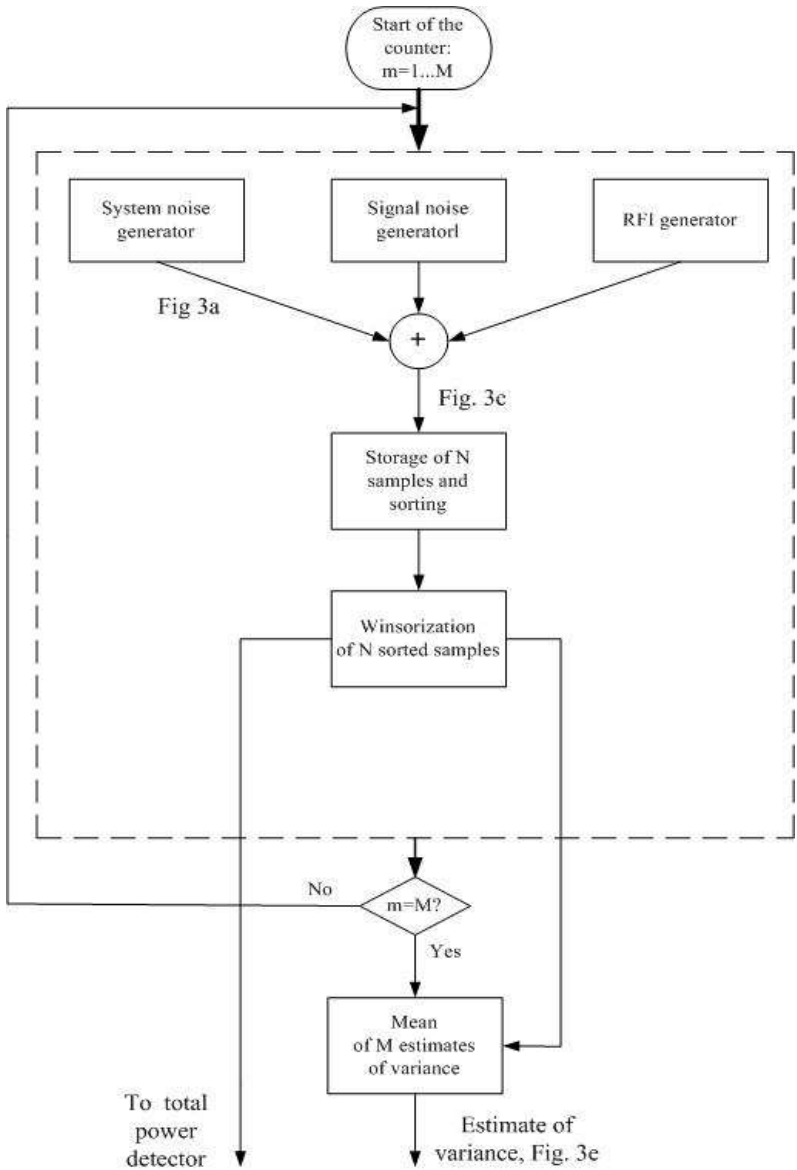}
\caption{Block diagram of computer simulations: winsorization is
used for RFI mitigation when  the total power detector is the
backend output as in section 4. 1.}
\end{figure}

\begin{figure}
\includegraphics[width=11.5cm,height=18.5cm]{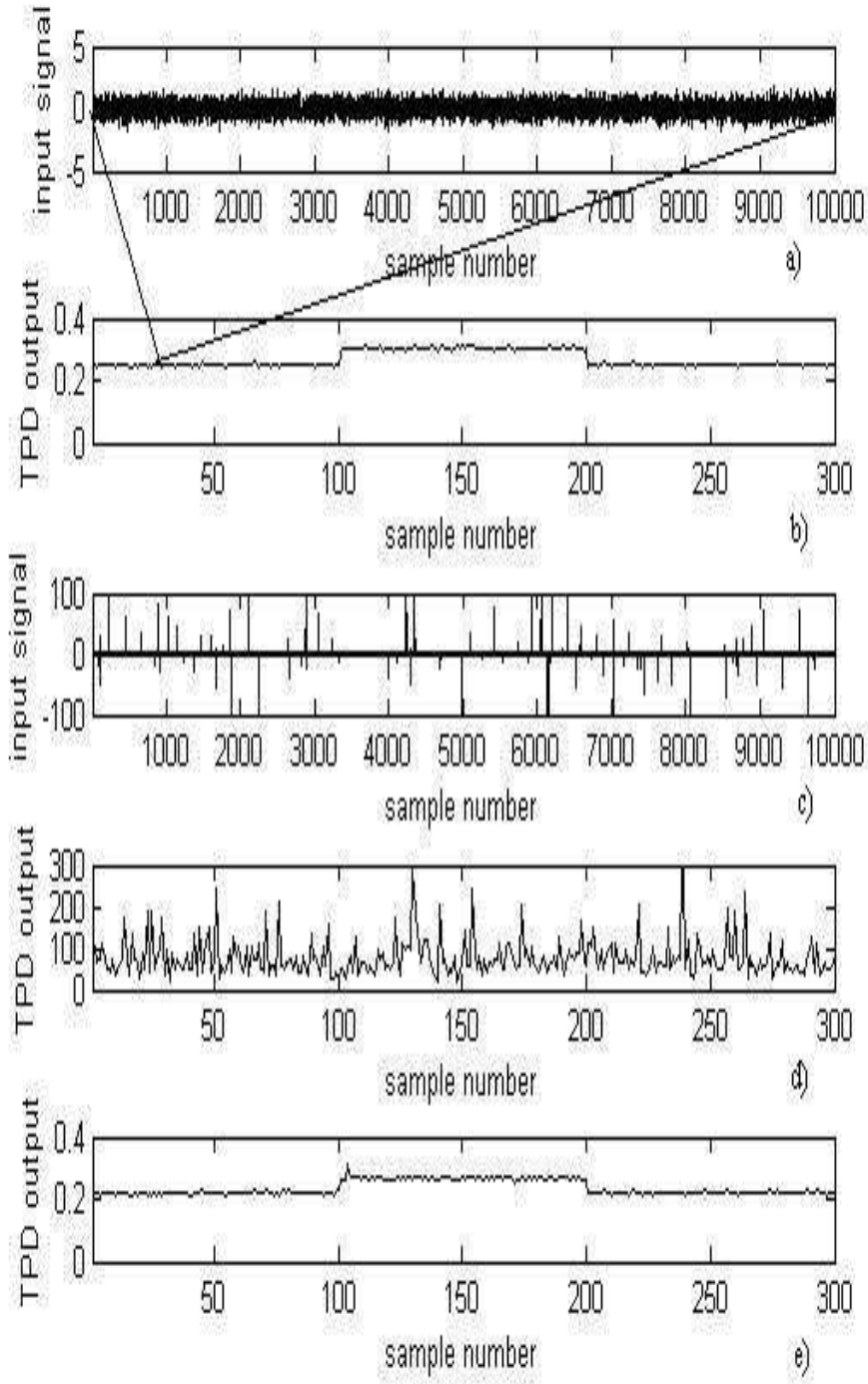}
\caption{Results of computer simulations with the algorithm shown in
Fig. 2: a) noise with the normal distribution, $\mu=0., \sigma=0.5$,
no interference; b) total power detector output, each point in this
figure corresponds to squaring and averaging of  $10^4$ samples in
figure a), there are two steps, ``up'' at point $\#100$ and
``down'' at point $\#200$  corresponding to the increase of $\sigma$
from value 0.5 to the value $\sigma+\Delta \sigma, \Delta
\sigma=0.05$. c) interference is added to the noise a): random
impulses with the Poisson distibution ($\lambda=0.004$) and the
lognormal disrtibution of amplitudes (mean=12, standard
deviation=6); d) total power detector output with  input signal  c);
e) total power detector output with  input signal c) and preliminary
winsorization,   note  the difference of scale in d) and e).}
\end{figure}

\begin{figure}
\includegraphics[width=11.5cm,height=18.5cm]{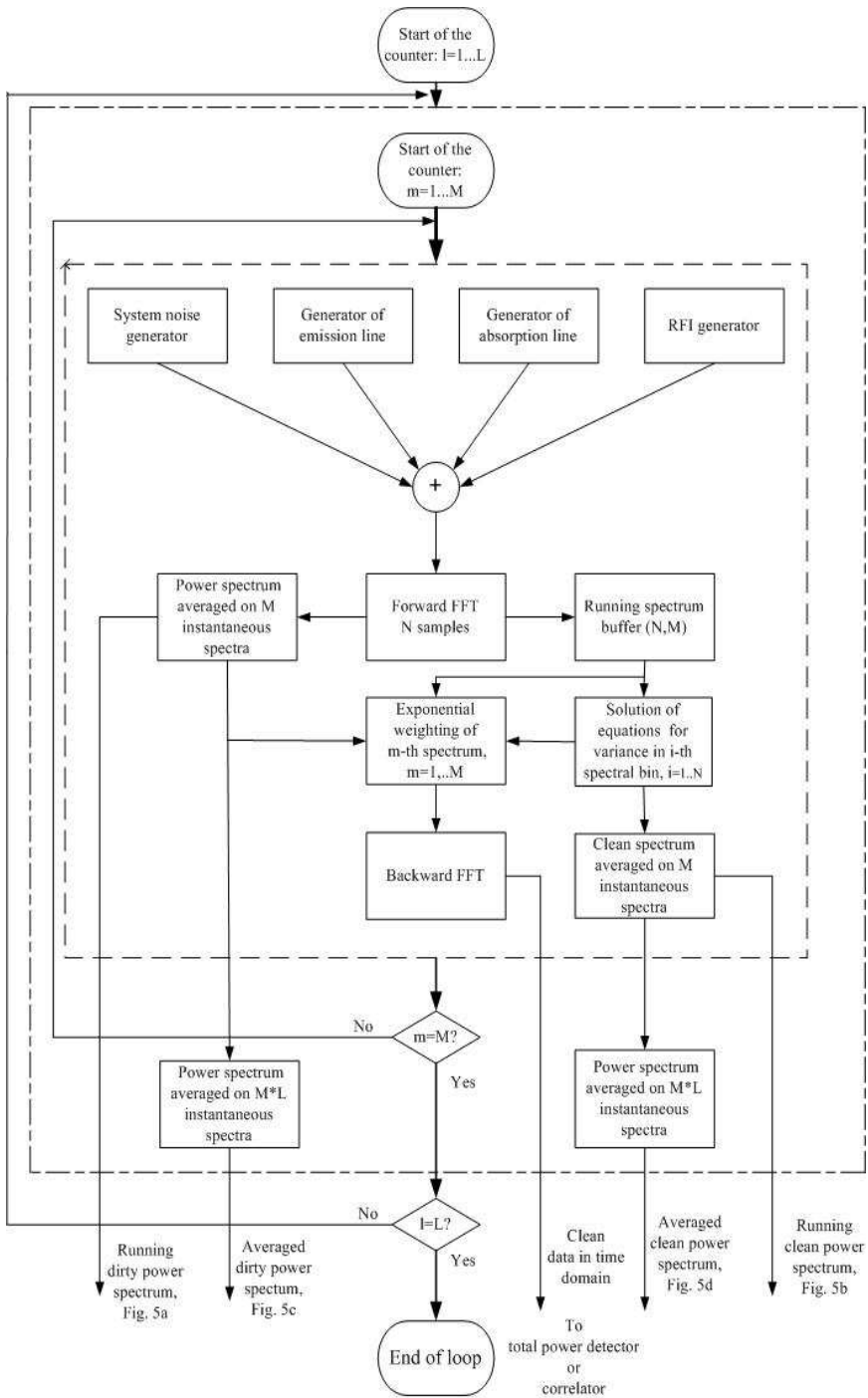}
\caption{Block diagram of  computer simulations: exponential
weighting is used for RFI mitigation  in the frequency domain when
the total power detector is the backend output as in section 4. 2
(pulsar observations)
 or the correlator as in section 4. 3.
(radiointerferometric observations).}
\end{figure}

\begin{figure}
\includegraphics[width=10.5cm,height=8.0cm]{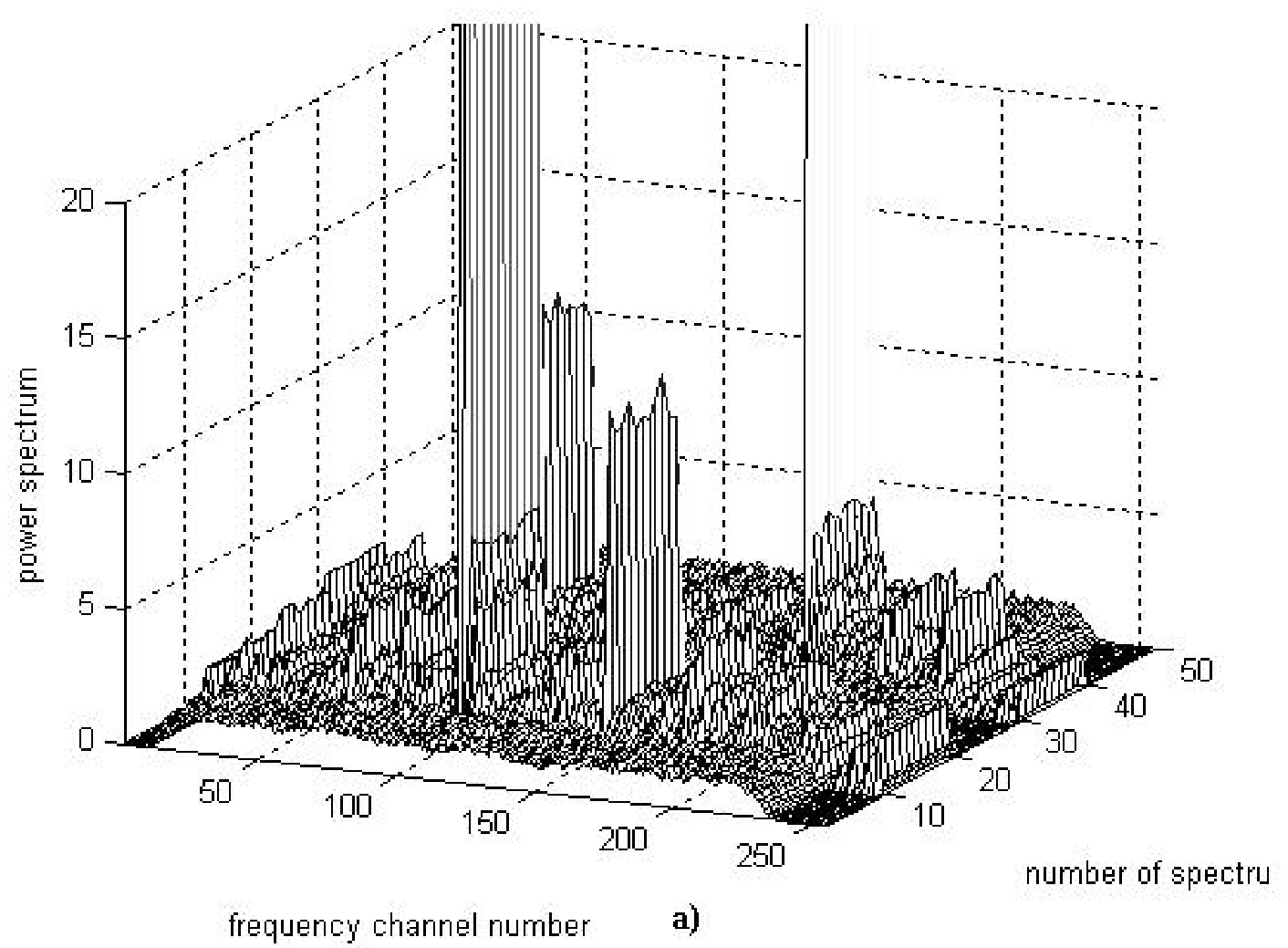}\\
\includegraphics[width=10.5cm,height=6.0cm]{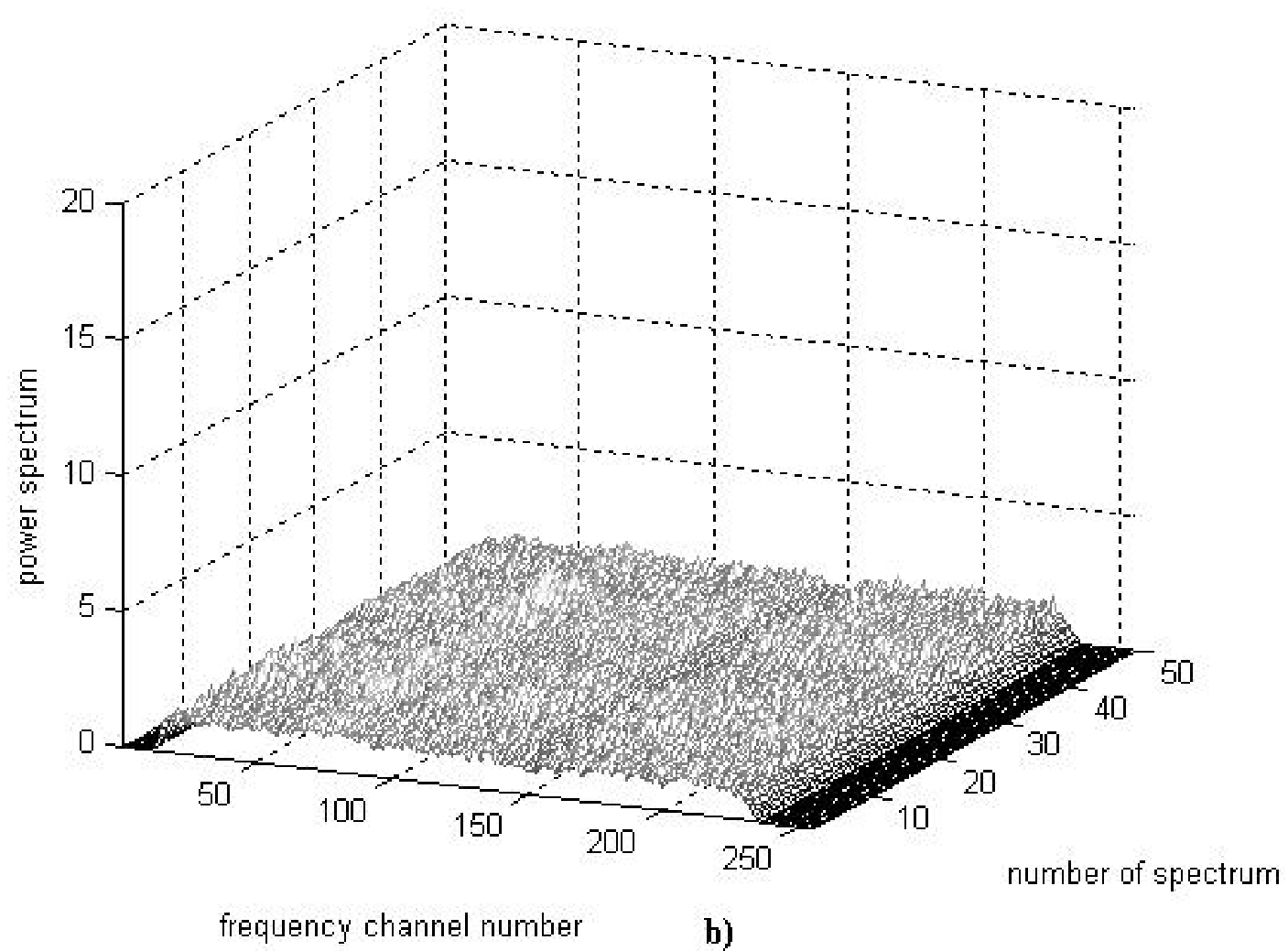}\\
\includegraphics[width=10.5cm,height=6.0cm]{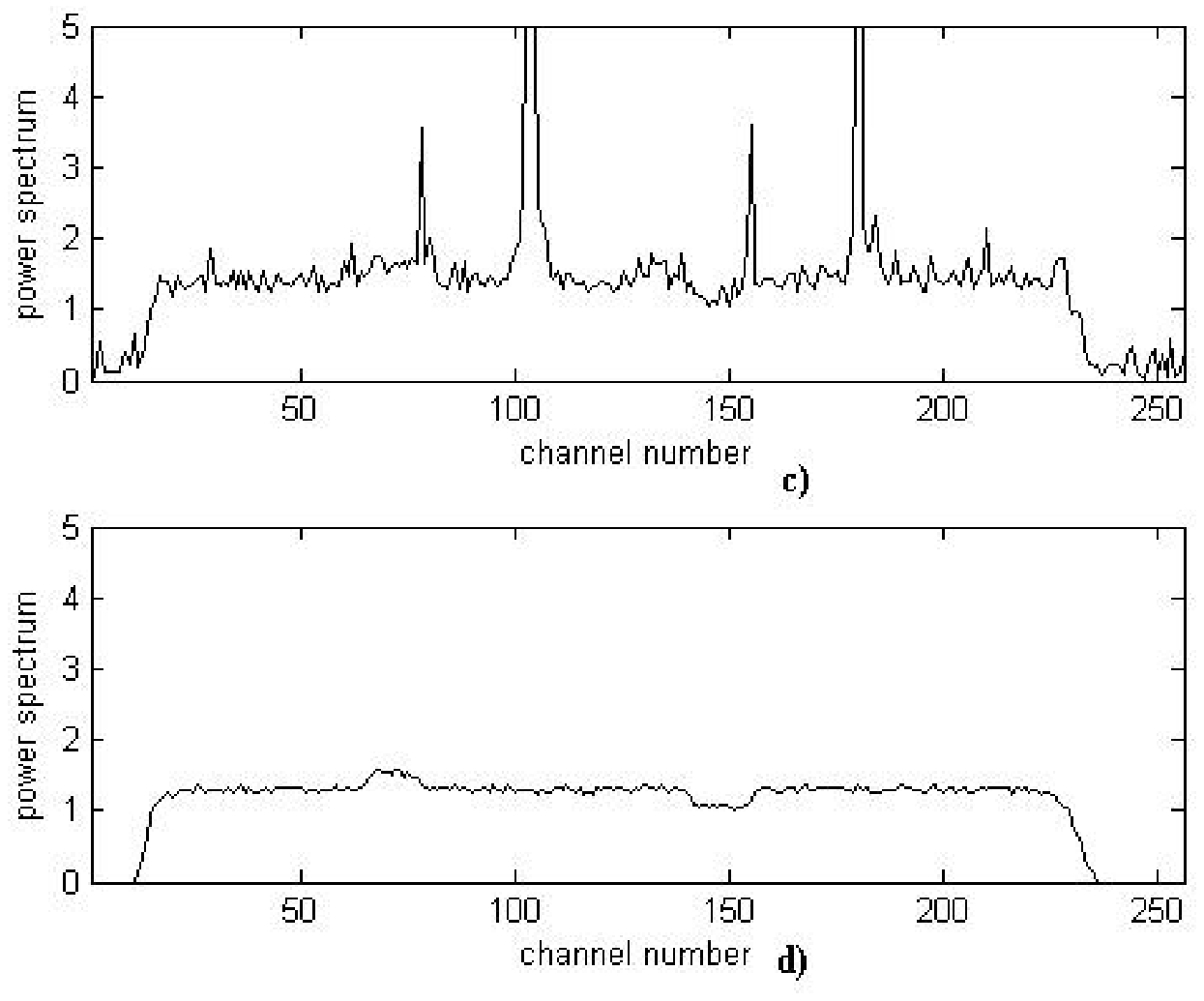}
\caption{Results of computer simulations  of exponential weighting
with the algorithm shown in Fig. 4: a) time-frequency presentation
of power spectrum consisting of system noise, emission and
absorption lines and RFI (randomly binary-phase manipulated
signals), L=50 time sections of spectrum divided on 256 channels,
each spectrum is the mean of $M=100$  instantaneous spectra; b)
power spectrum obtained as a solution to equation (18) for each
spectral channel and $M=100$  samples; c) power spectrum  averaged
in time, using $L=50$ sample spectra from a); d) power spectrum
averaged in time, using $L=50$ sample spectra from b), i.e., with
RFI mitigation. }
\end{figure}

\begin{figure}
\includegraphics[width=10.5cm,height=6.0cm]{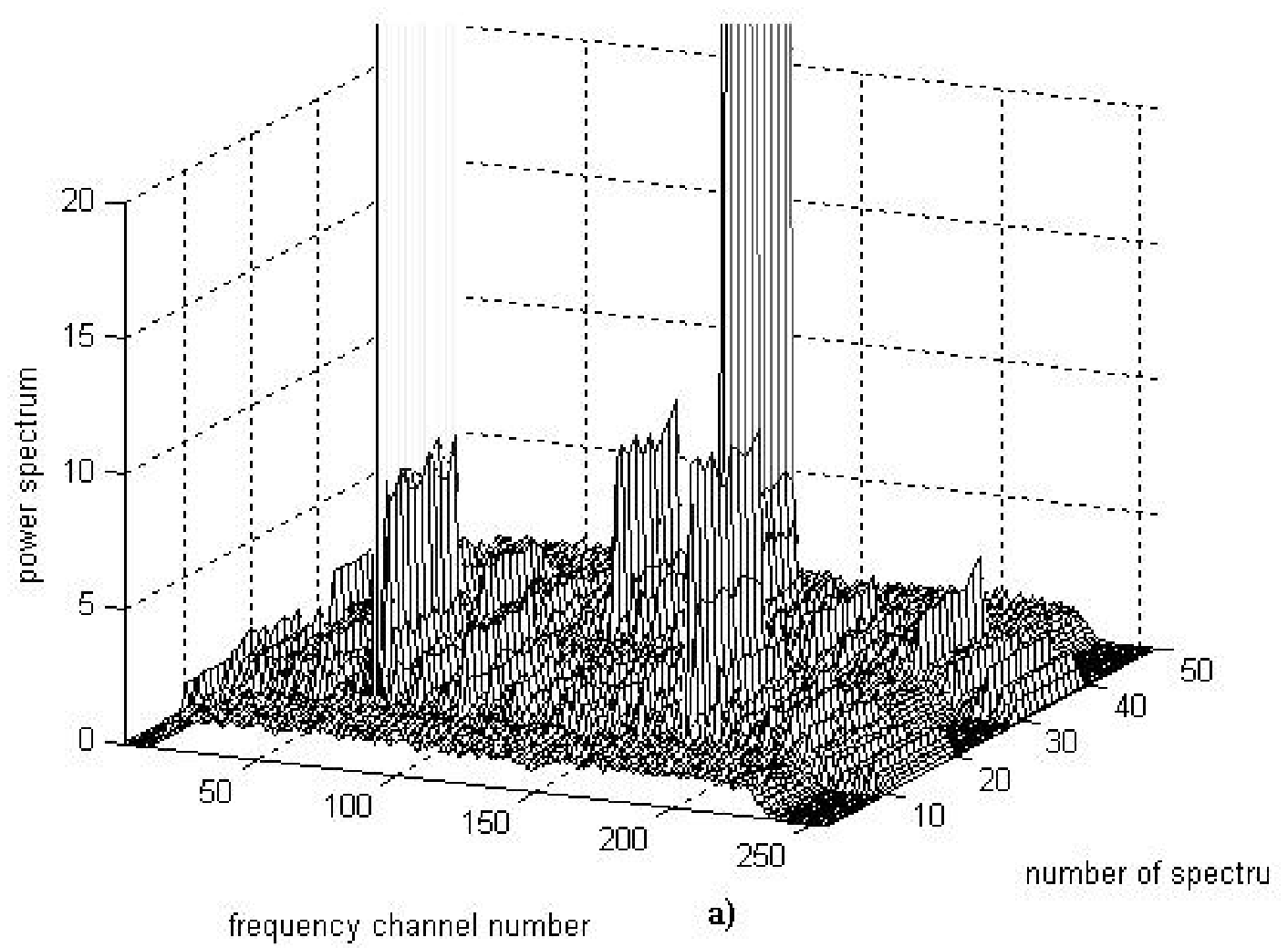}\\
\includegraphics[width=10.5cm,height=6.0cm]{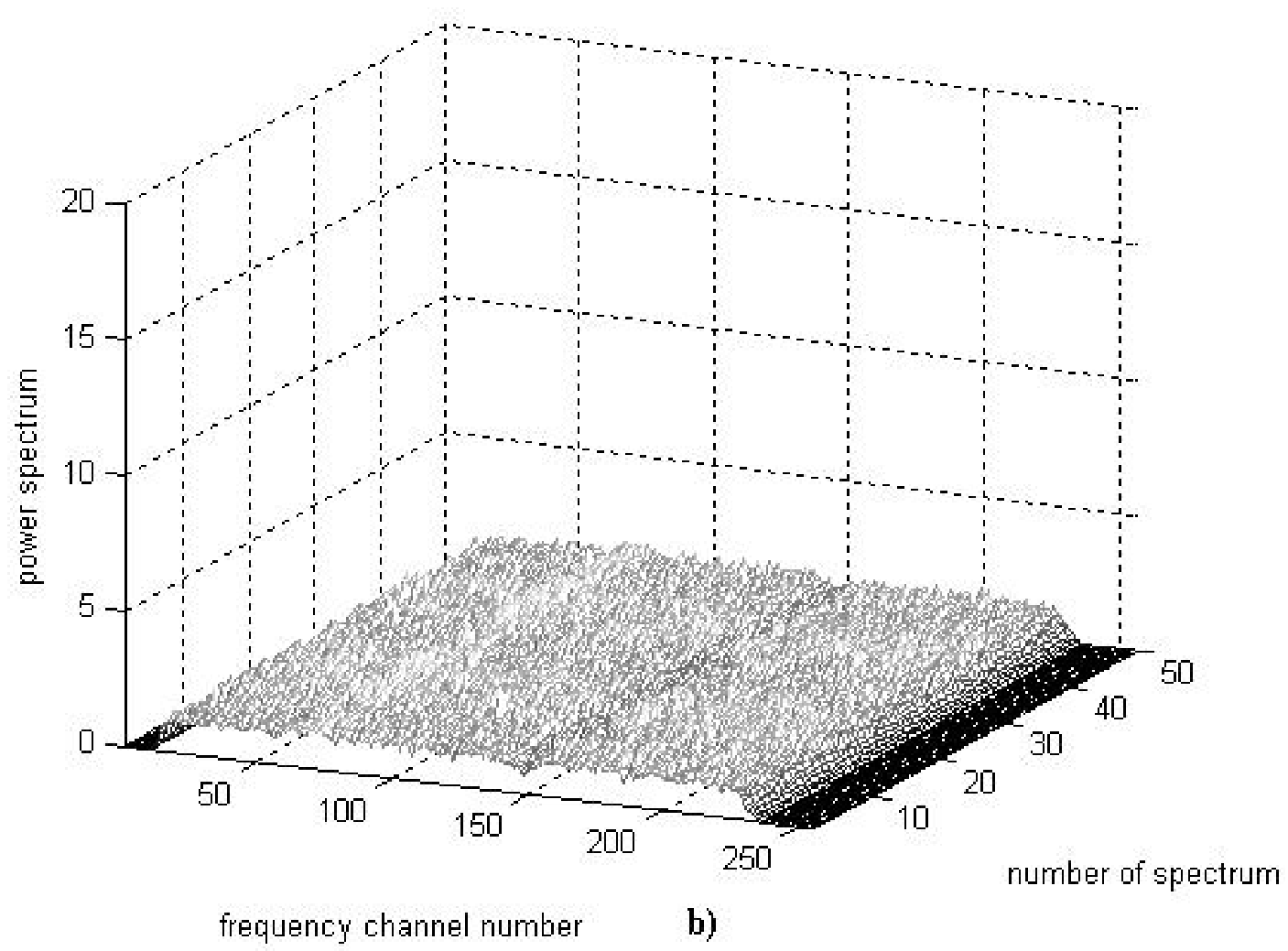}\\
\includegraphics[width=10.5cm,height=8.0cm]{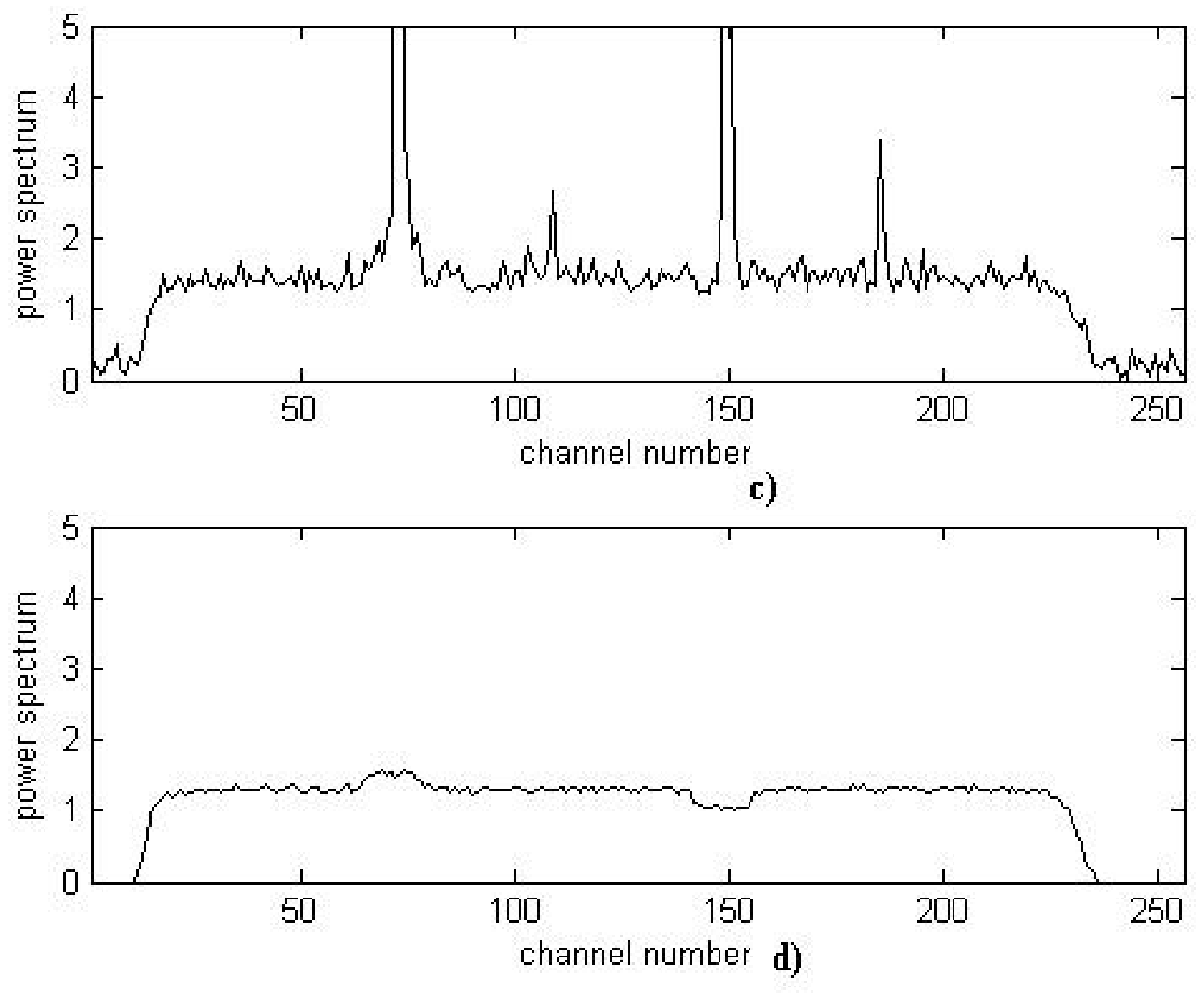}
\caption{Results of computer simulations  of exponential weighting
with the algorithm shown in Fig. 4: spectra  of interference
overlap with  spectral lines. Other parameters are similar to Fig.
5, a) time-frequency presentation of the power spectrum consisting
of system noise, emission and absorption lines and RFI; b) power
spectrum obtained as a solution to equation (18) for each spectral
channel and $M=100$  samples; c) power spectrum  averaged in time,
using $L=50$ sample spectra from a); d) power spectrum  averaged in
time, using $L=50$ sample spectra from b), i.e., with RFI
mitigation. Both spectral lines are clearly visible.}
\end{figure}

\begin{figure}
\includegraphics[width=10.5cm,height=8.0cm]{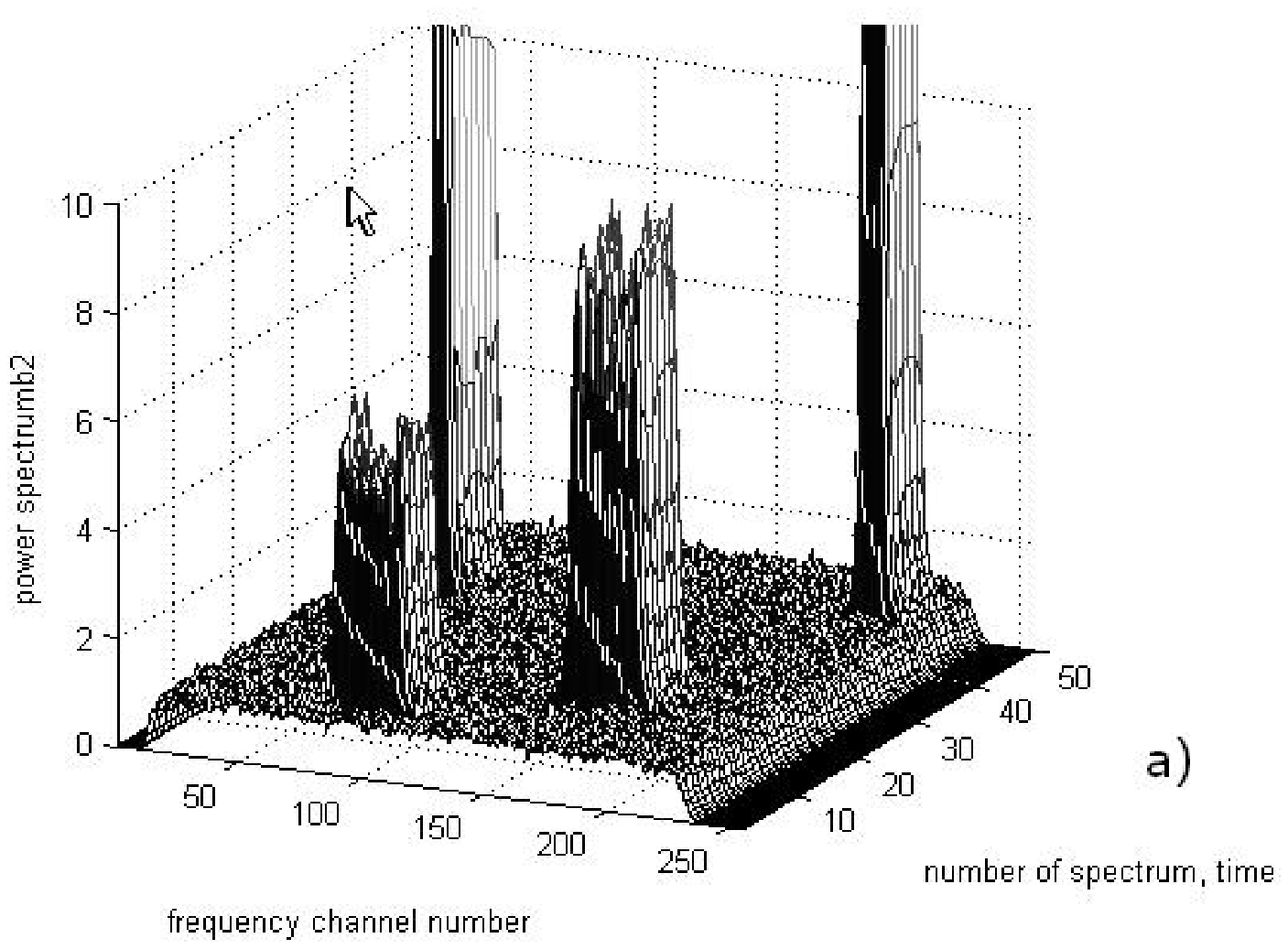}\\
\includegraphics[width=10.5cm,height=5.0cm]{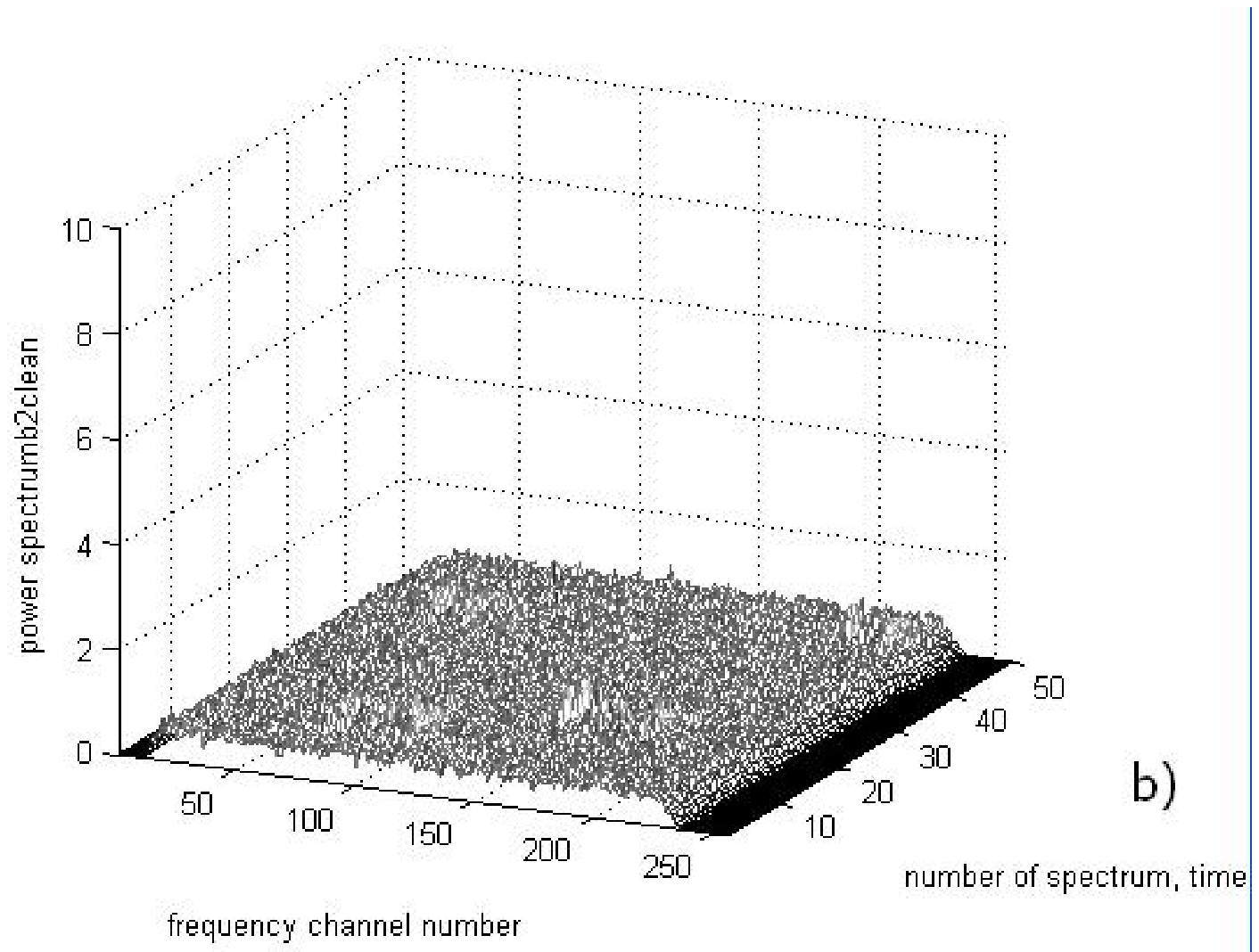}\\
\includegraphics[width=10.5cm,height=5.0cm]{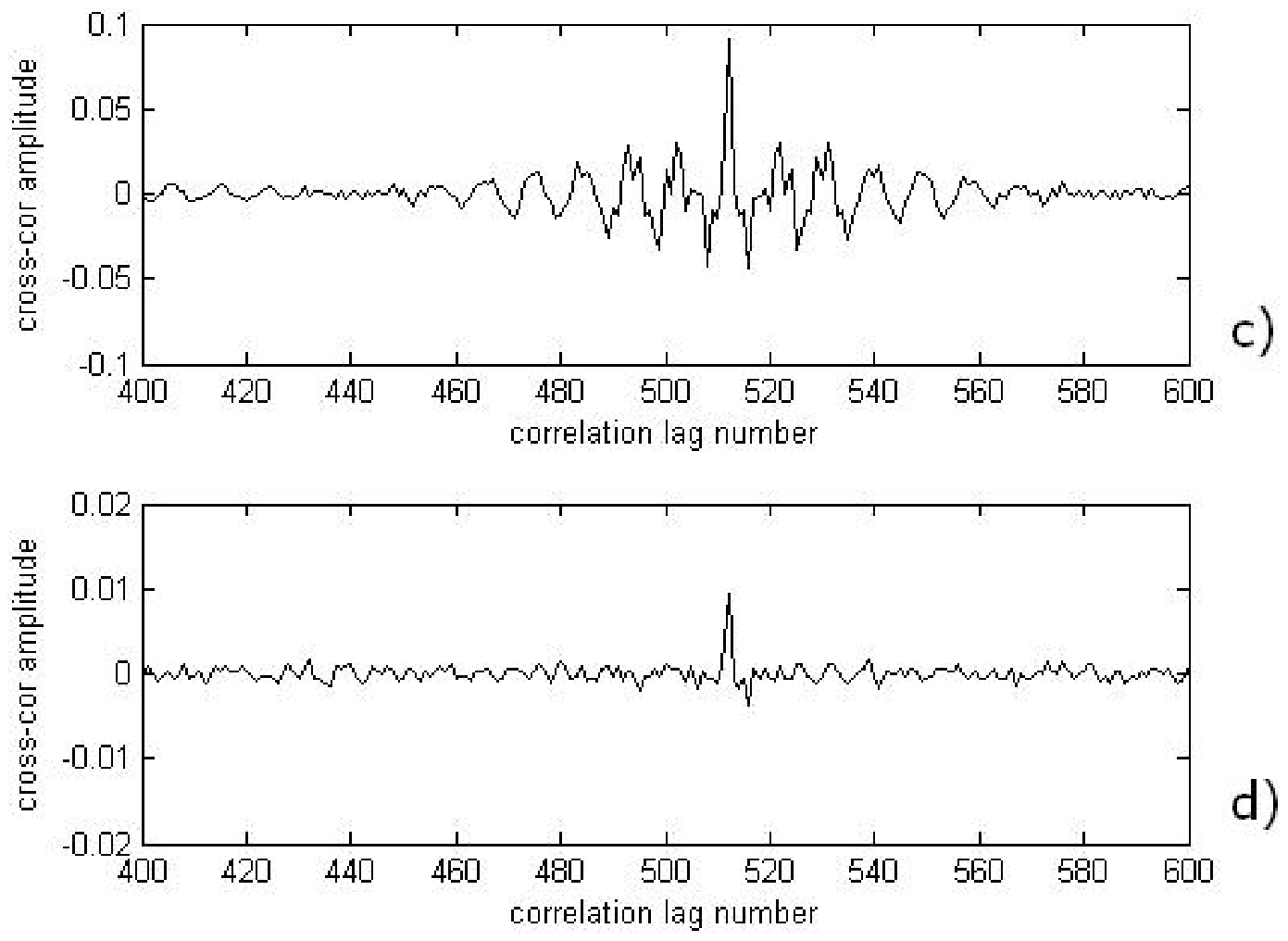}
\caption{Results of computer simulations  of exponential weighting
with the algorithm shown in Fig. 4 when ``dirty'' and ``clean''
signals are applied to the correlator; L=50 time sections of the
spectrum divided on 256 spectral channels, each spectrum is the mean
of $M=100$  instantaneous spectra; the spectra at the first input of
the correlator are shown, the  spectra at the second  input are
similar to a) and b): a) time-frequency presentation of the power
spectrum consisting of system noise, and RFI (frequency-modulated
bursts); b) power spectrum after RFI mitigation - exponential
weighting of each instantaneous spectrum using variances obtained as
a solution to equation (18) for each spectral channel and $M=100$
samples; c) cross-correlation in the presence of RFI and no RFI
mitigation, the central 200 channels are shown; d) cross-correlation
in the presence of RFI and with RFI mitigation; take notice of the
difference of the vertical scales in c) and d).}
\end{figure}

\begin{figure}
\includegraphics[width=9.0cm,height=9.0cm]{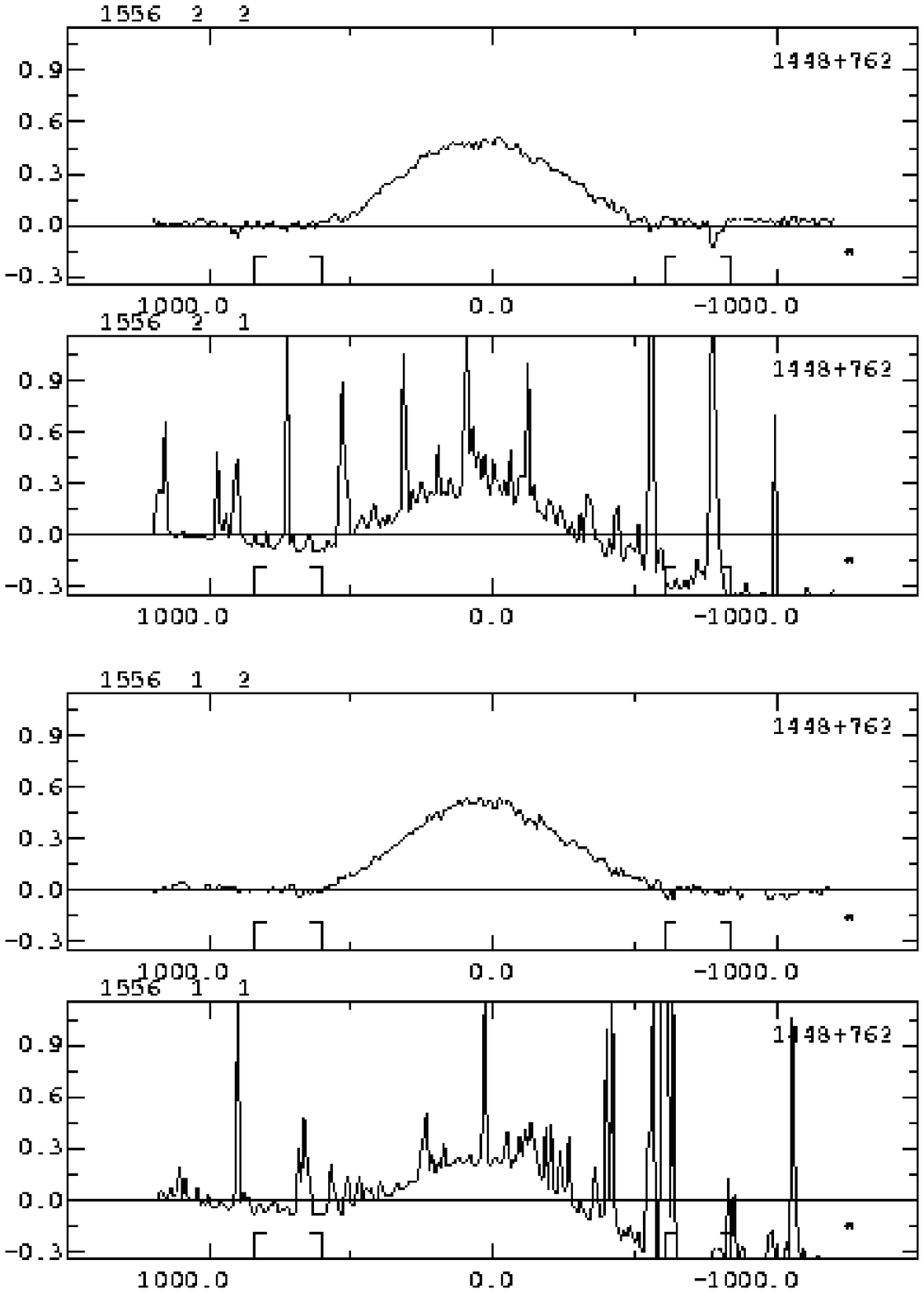}
\includegraphics[width=9.0cm,height=9.0cm]{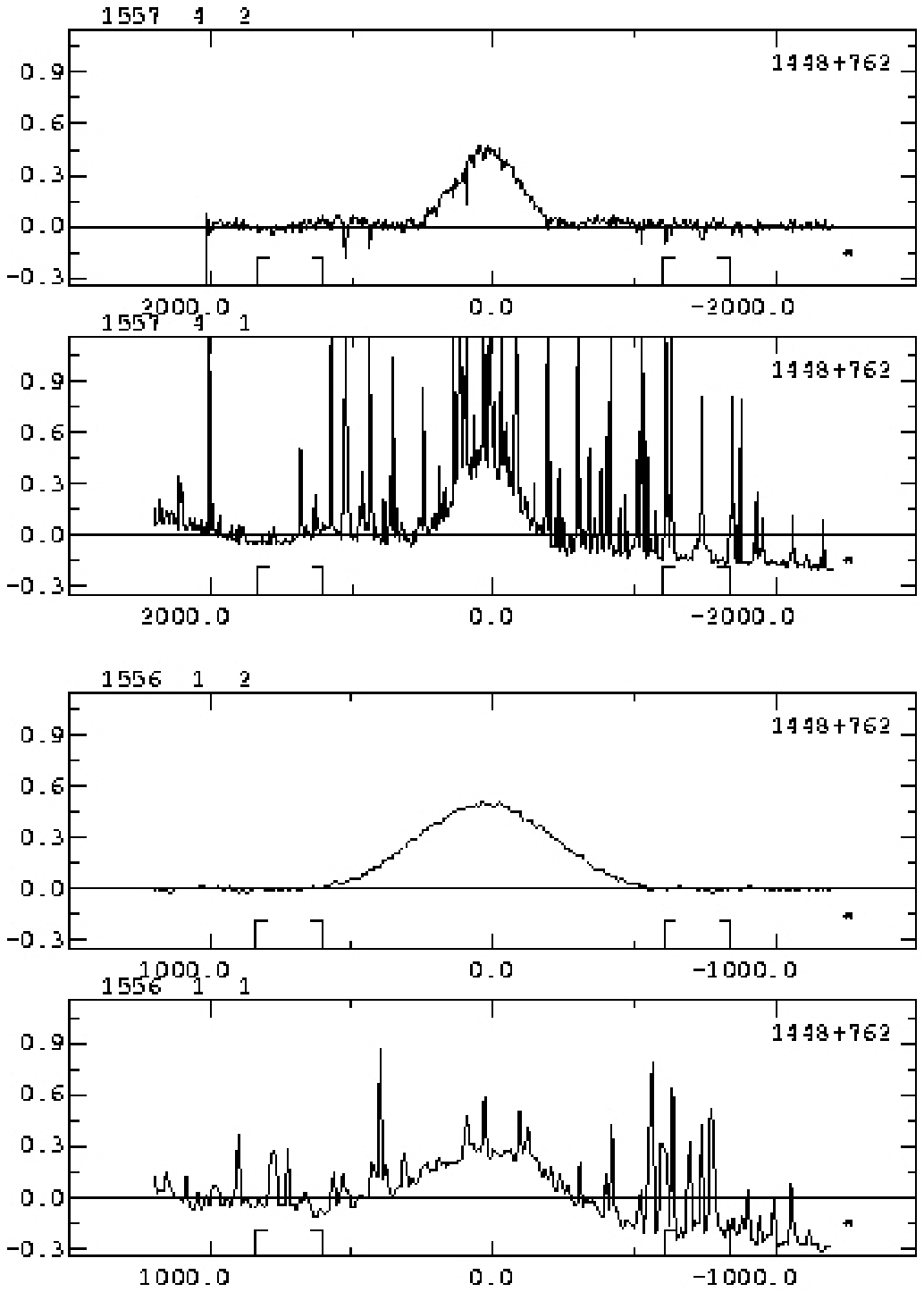}
\end{figure}

\begin{figure}
\includegraphics[width=9.0cm,height=9.0cm]{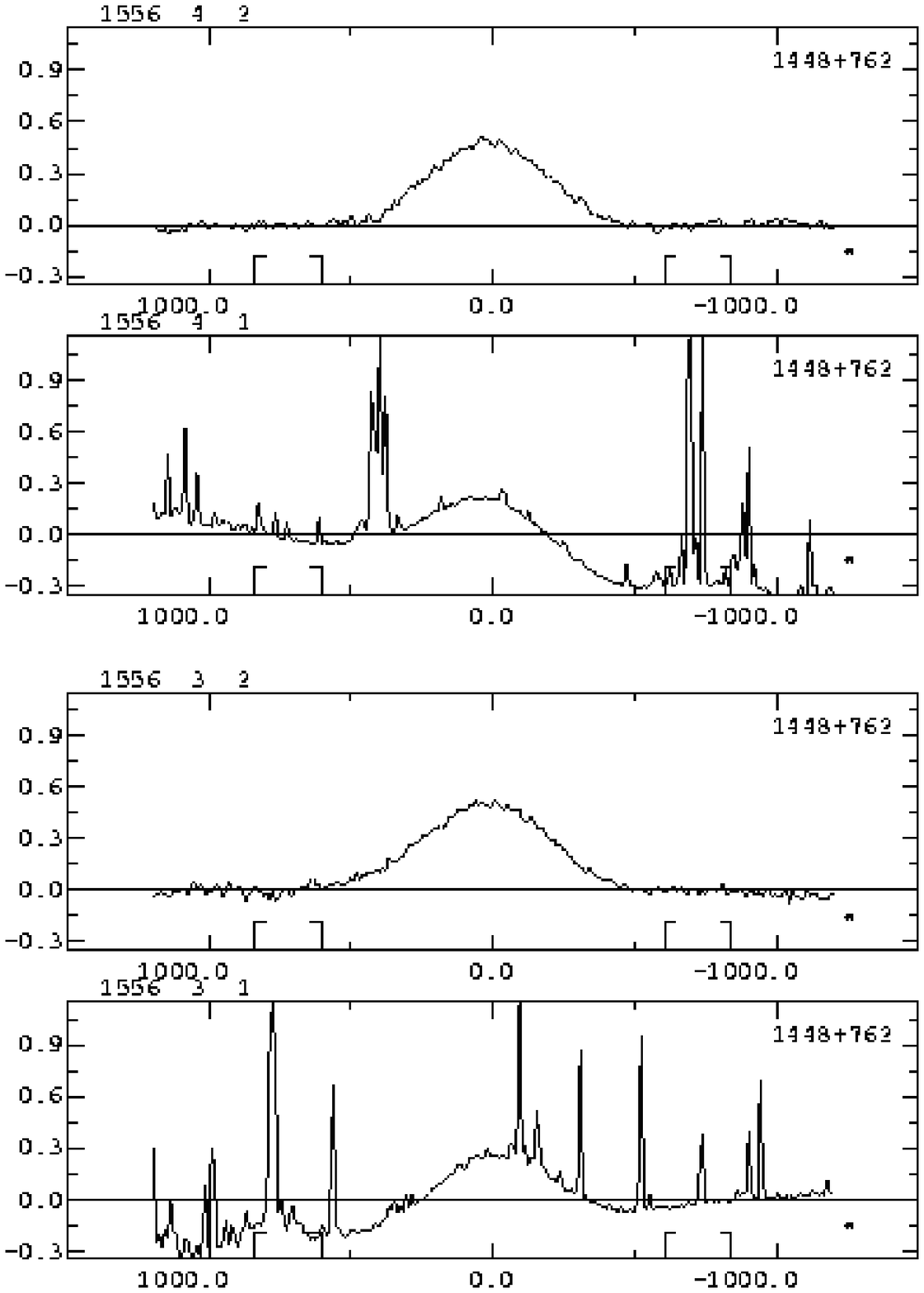}
\includegraphics[width=9.0cm,height=9.0cm]{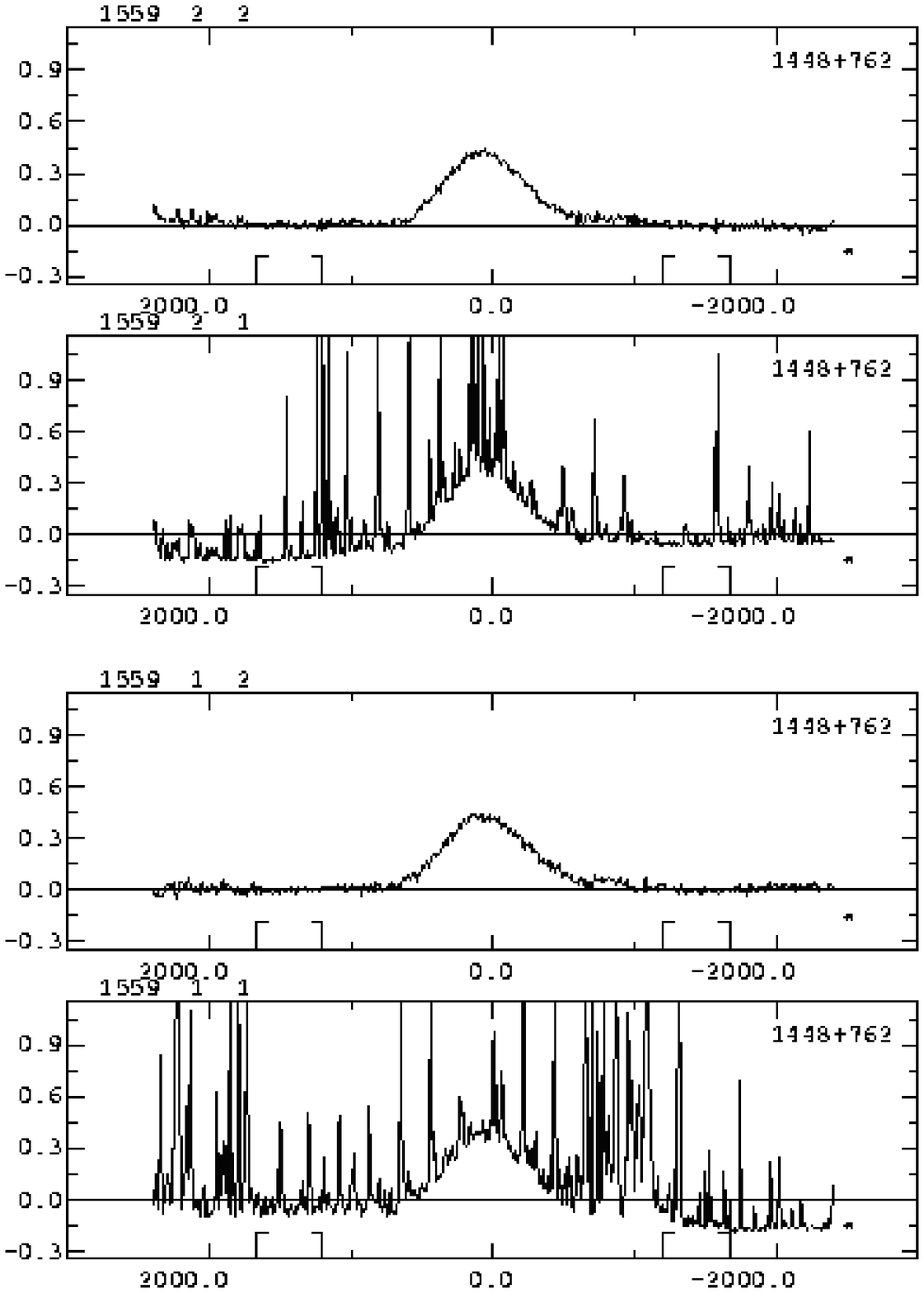}
\caption{Examples  of  RFI mitigation at the Effelsberg radio
telescope during observations in continuum at central frequency 1645
MHz, bandwidth 20MHz. A selection of  eight scans  of the source
1448+762 is represented. The pairwise records were made
simultaneously for the channel with RFI  mitigation  and the channel
without RFI mitigation.}
\end{figure}

\begin{figure}
\includegraphics[width=9.0cm,height=9.0cm]{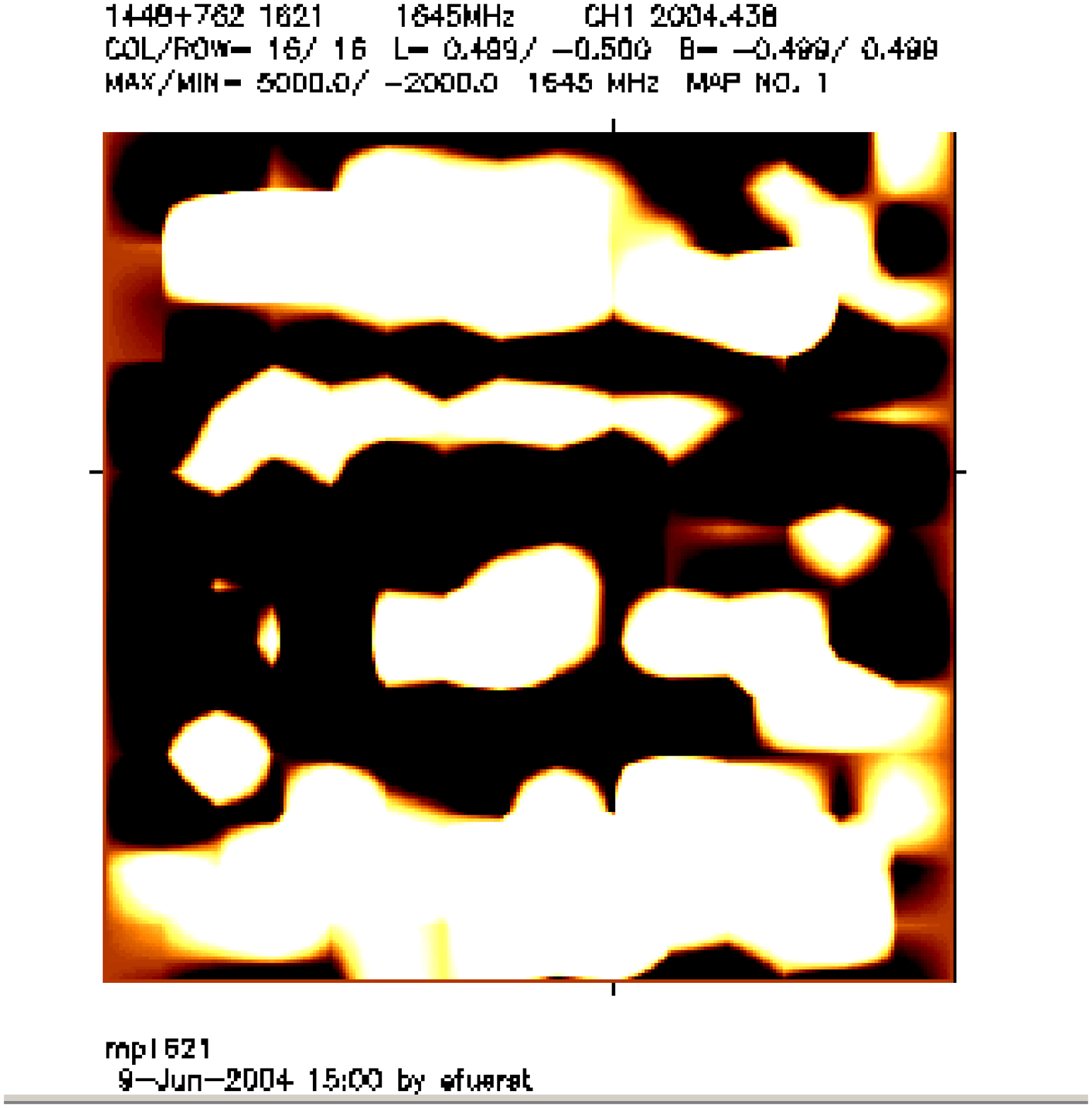}
\includegraphics[width=9.0cm,height=9.0cm]{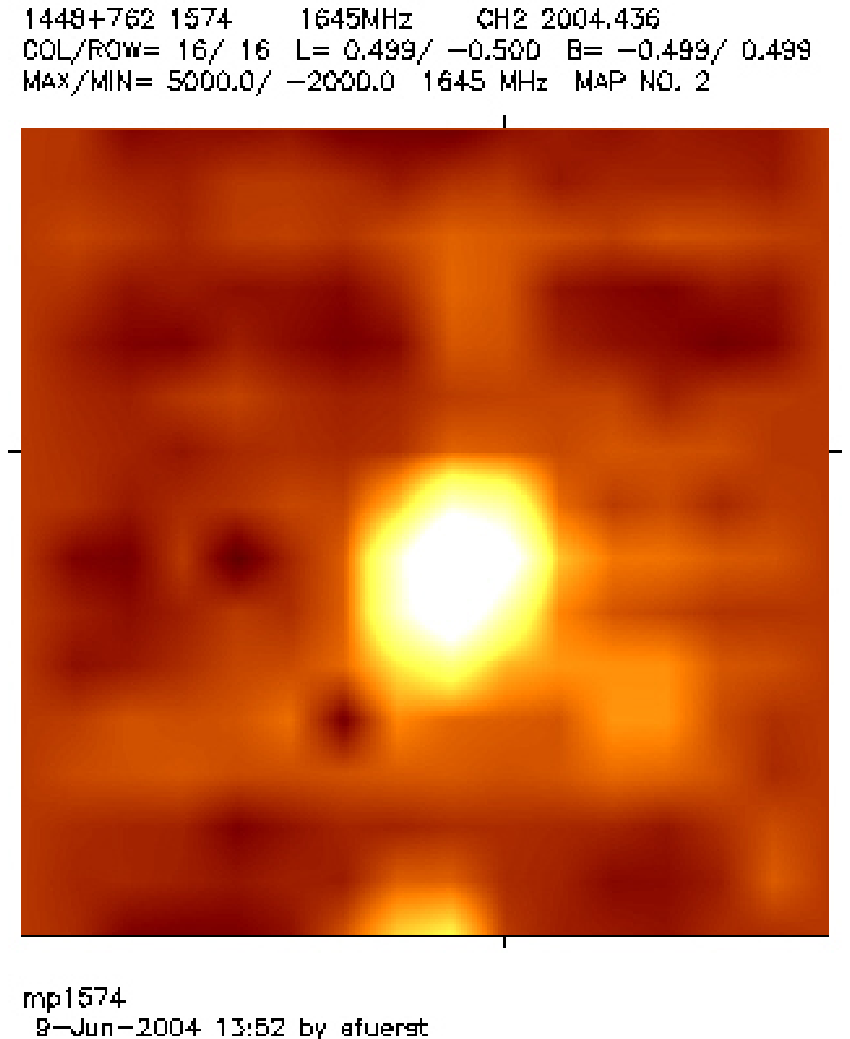}
\caption{Radio image of the source 1448+762 built using  scans
similar to those in Fig. 2:  left panel - without RFI mitigation,
right panel - with RFI mitigation.}

\end{figure}

\clearpage

\begin{figure}
\includegraphics[width=9.0cm,height=6.0cm]{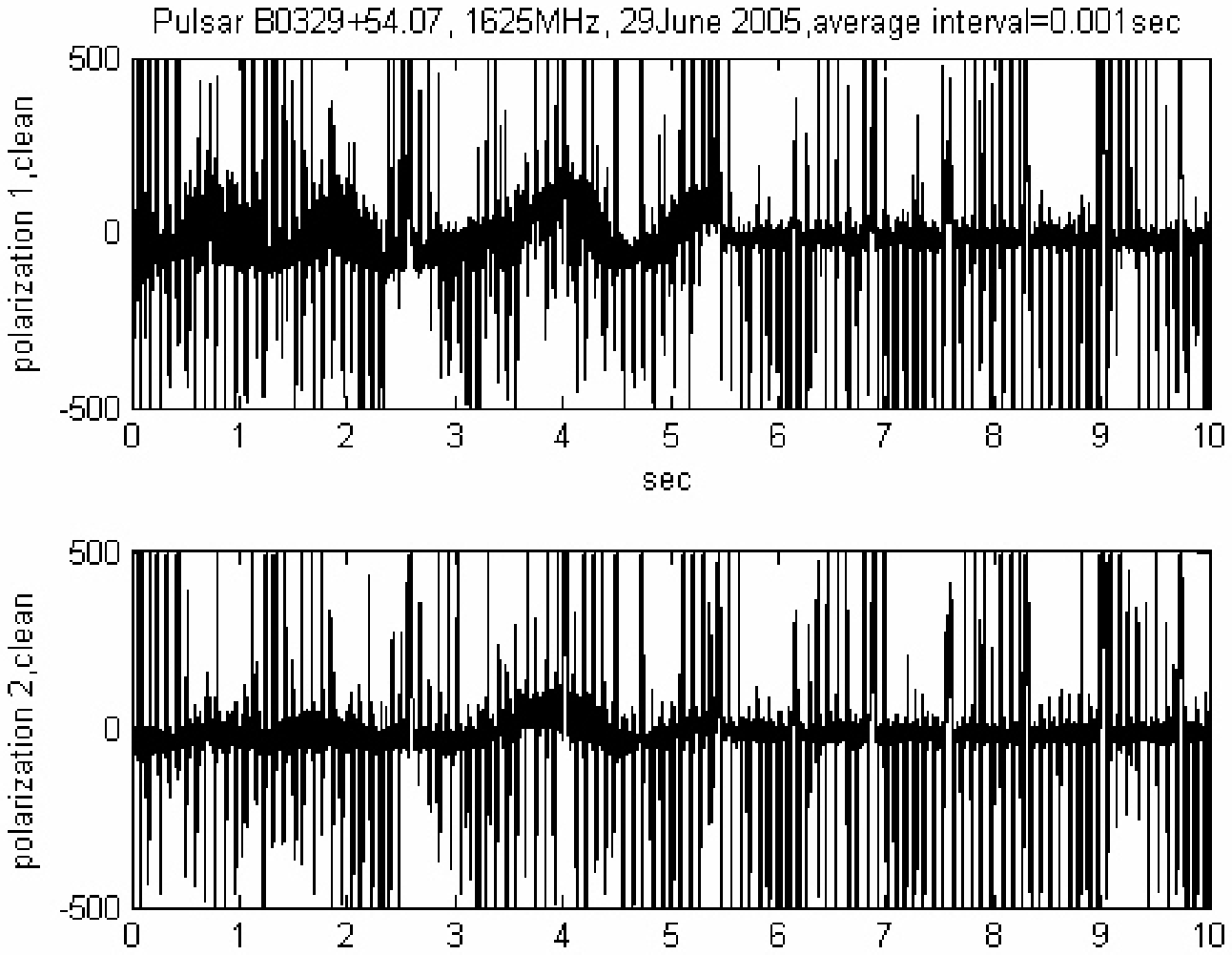}
\includegraphics[width=9.0cm,height=6.0cm]{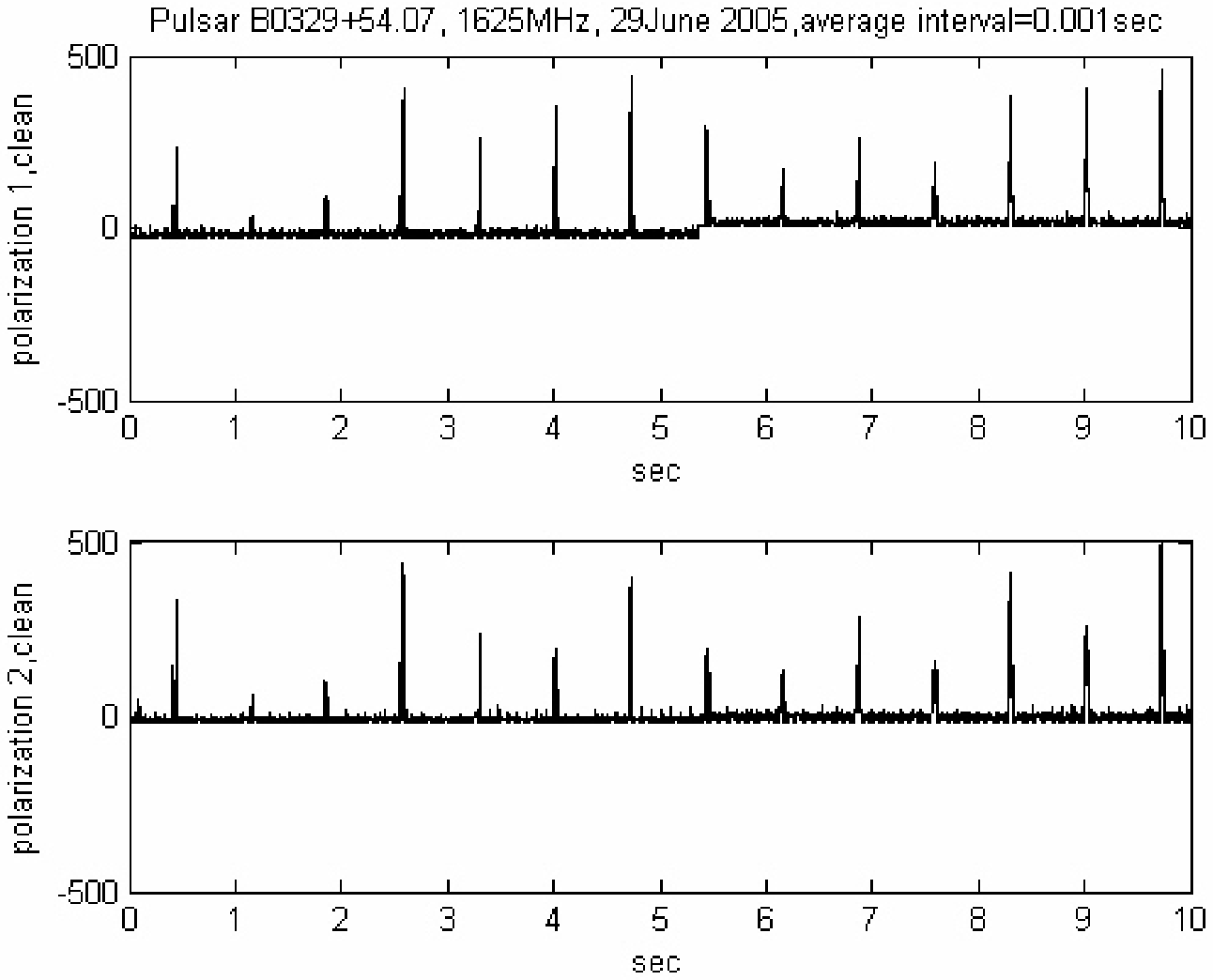}
\end{figure}

\begin{figure}
\includegraphics[width=9.0cm,height=6.0cm]{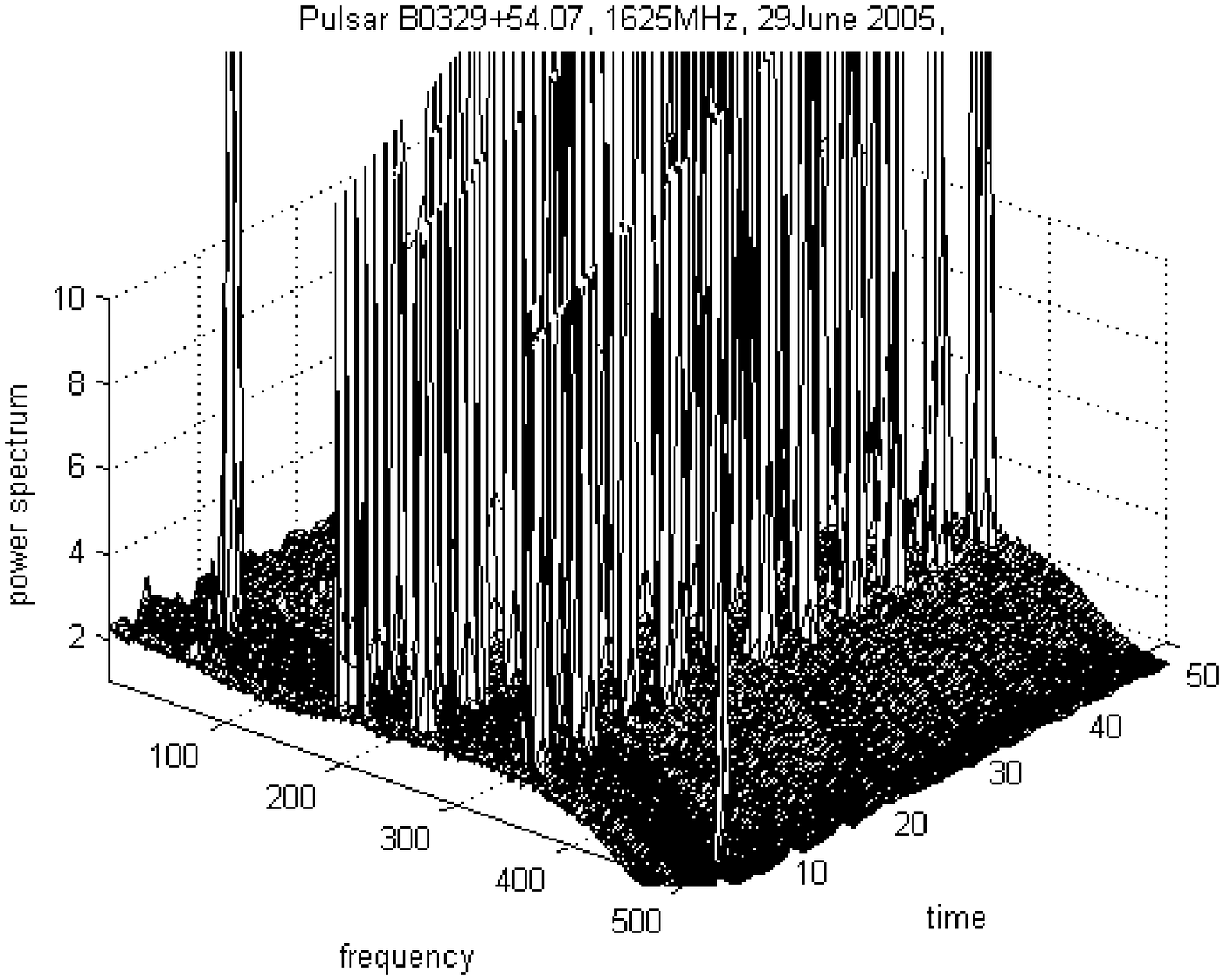}
\includegraphics[width=9.0cm,height=6.0cm]{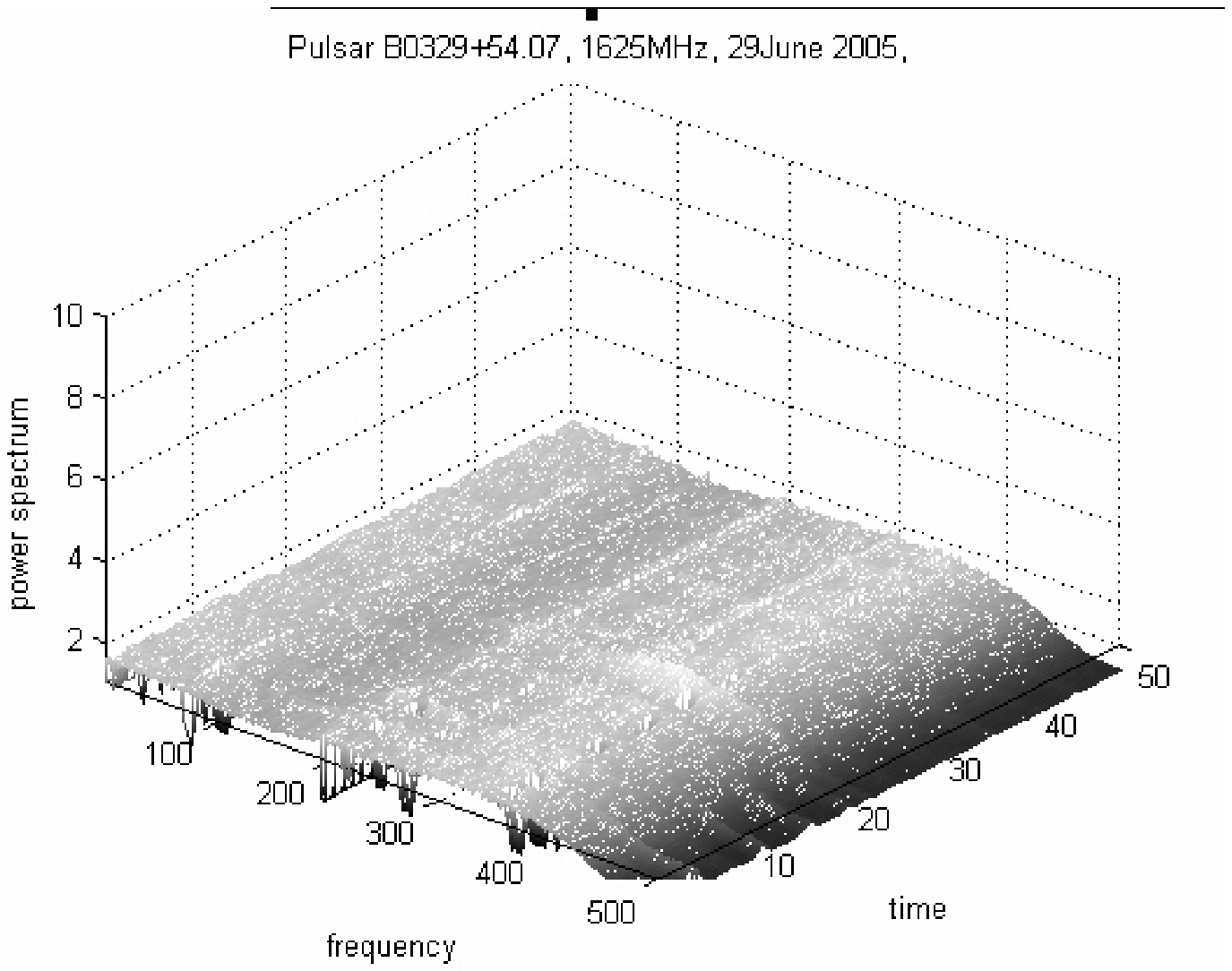}
\end{figure}

\begin{figure}
\includegraphics[width=5.5cm,height=5.5cm]{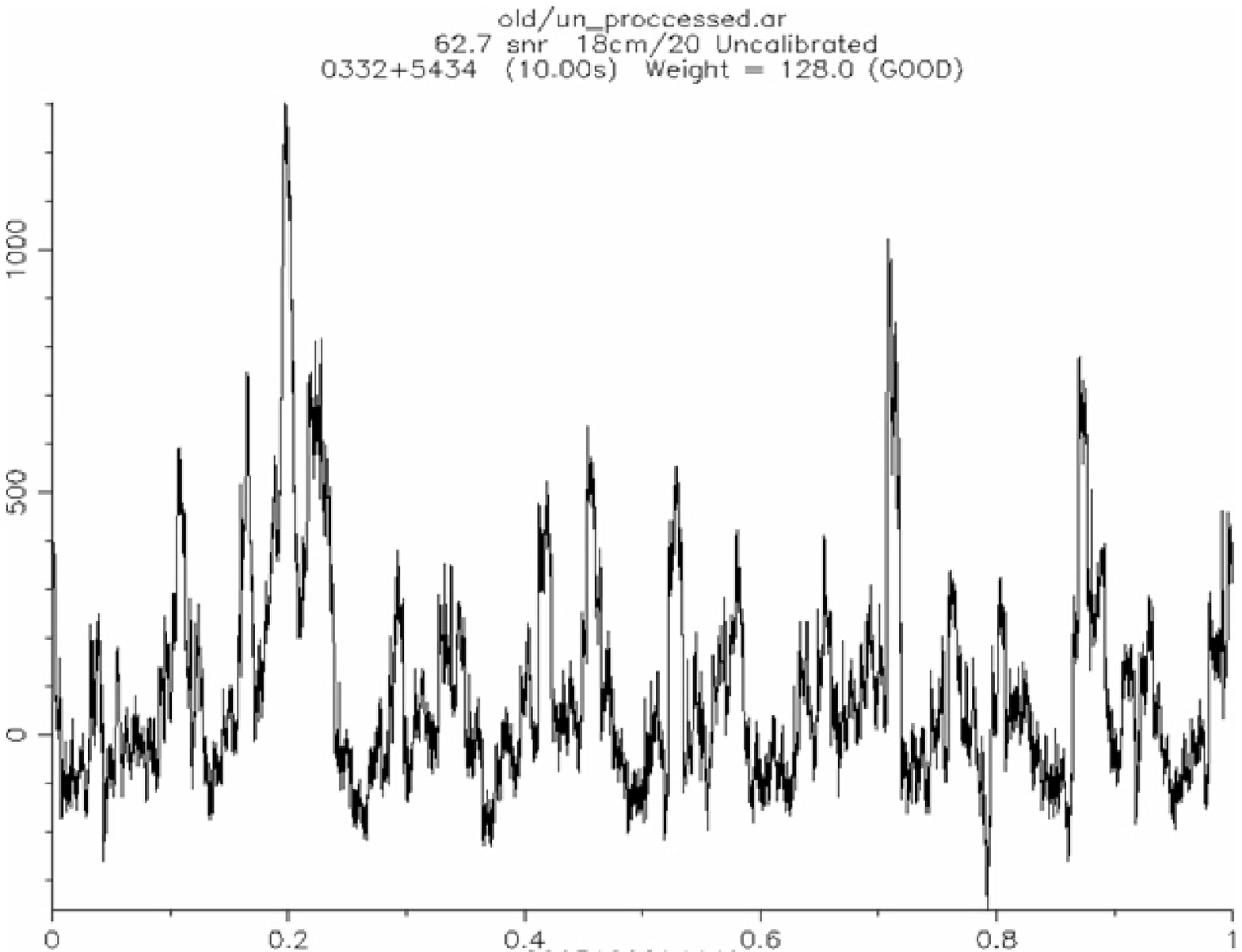}
\includegraphics[width=5.5cm,height=5.5cm]{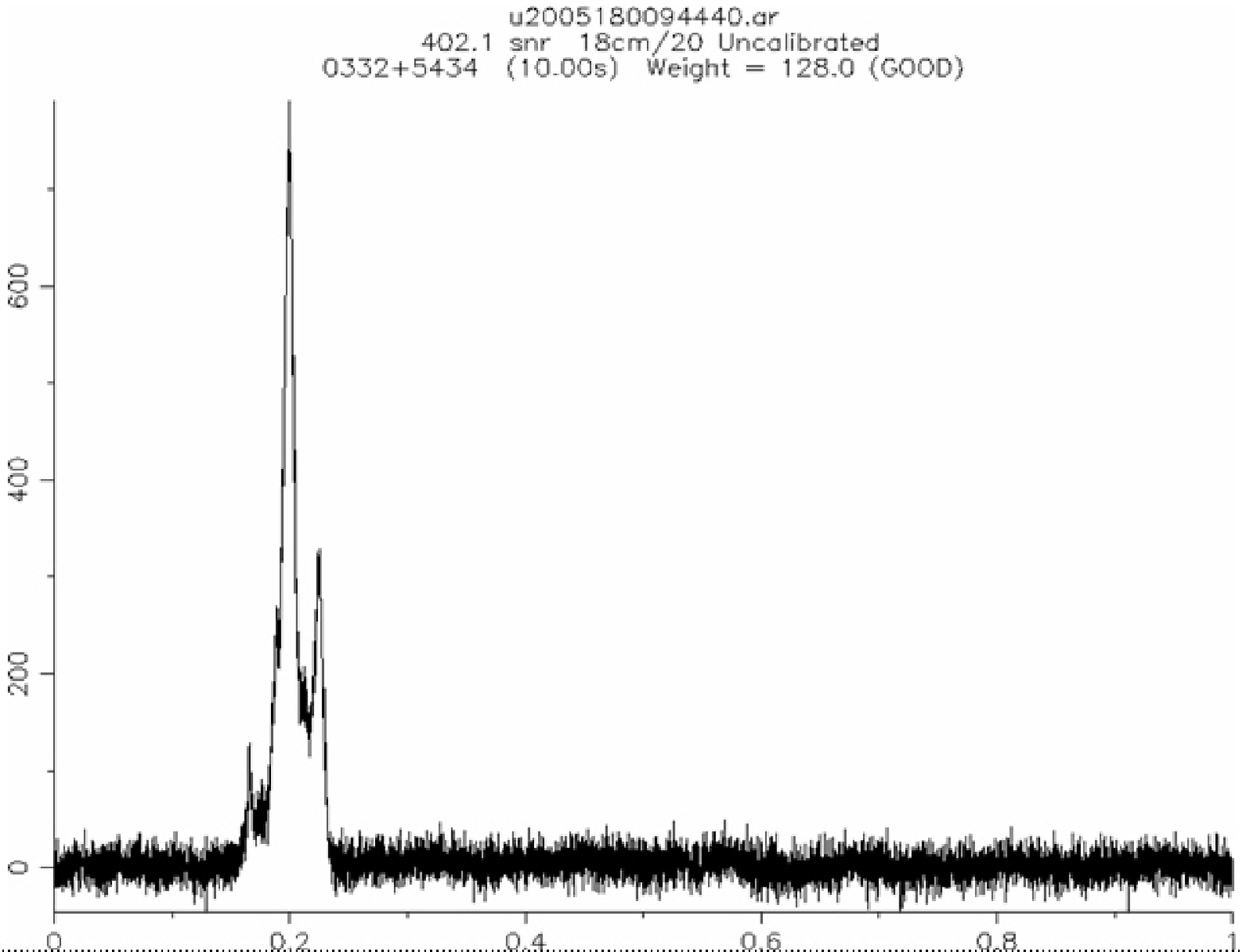}
\includegraphics[width=5.5cm,height=5.5cm]{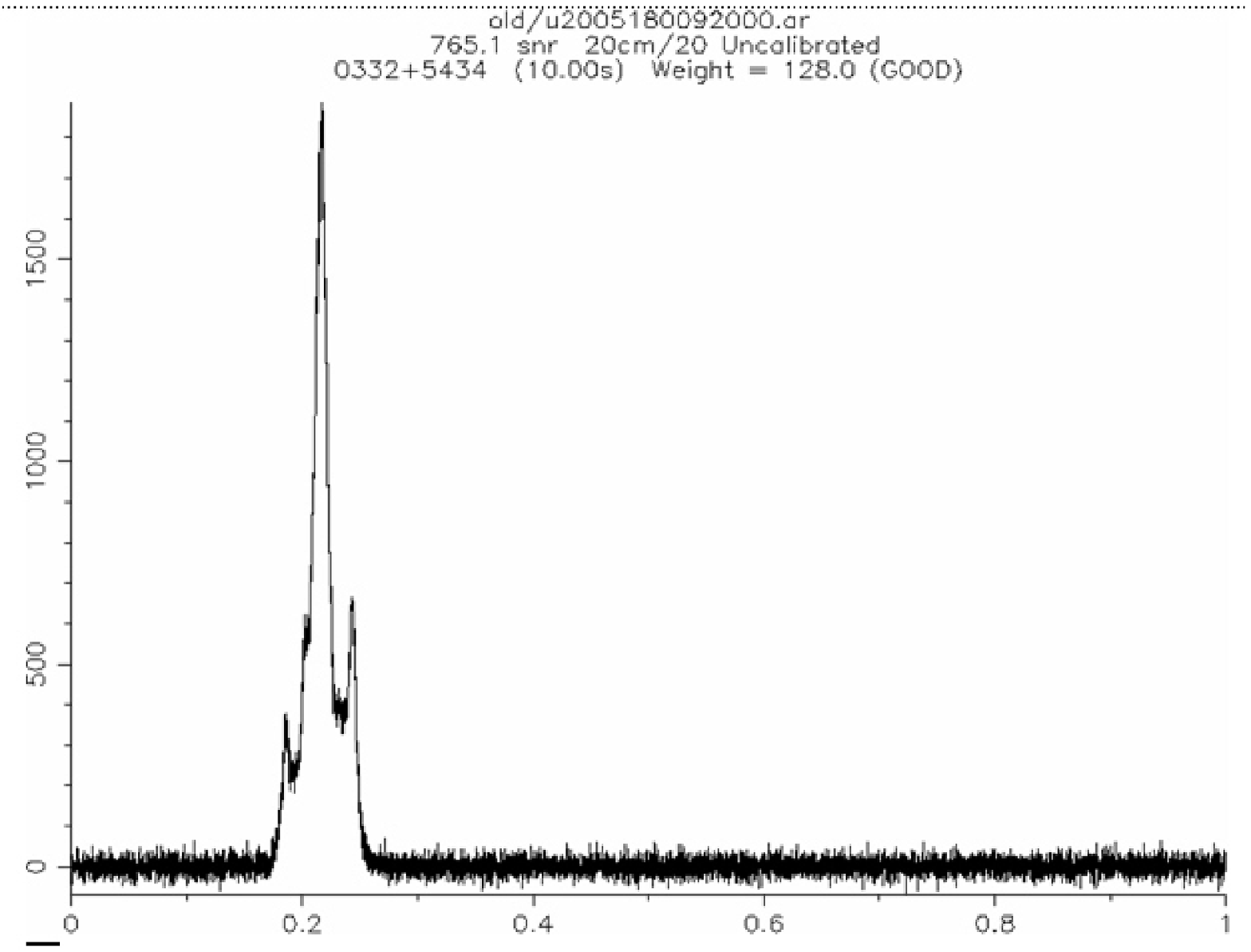}

\caption{ Pulsar B0329+54.07 observed at WSRT at 1625MHz. Data were
recorded during 10 sec, 40 Msamples/sec. Upper row, left panel: TPD
ouputs  for two polarizations, raw data with RFI, right panel: TPD
outputs, RFI removed. Middle row, right panel: time fragment  of the
running power spectrum with RFI; the same time fragment, RFI
removed. Lower row, left panel: pulsar profile made with raw data
over 10 sec, middle panel: pulsar profile, RFI removed, right panel:
pulsar observed at 1420 MHz, no RFI.}

\end{figure}

\clearpage
\begin{figure}
\includegraphics[width=9.0cm,height=9.0cm]{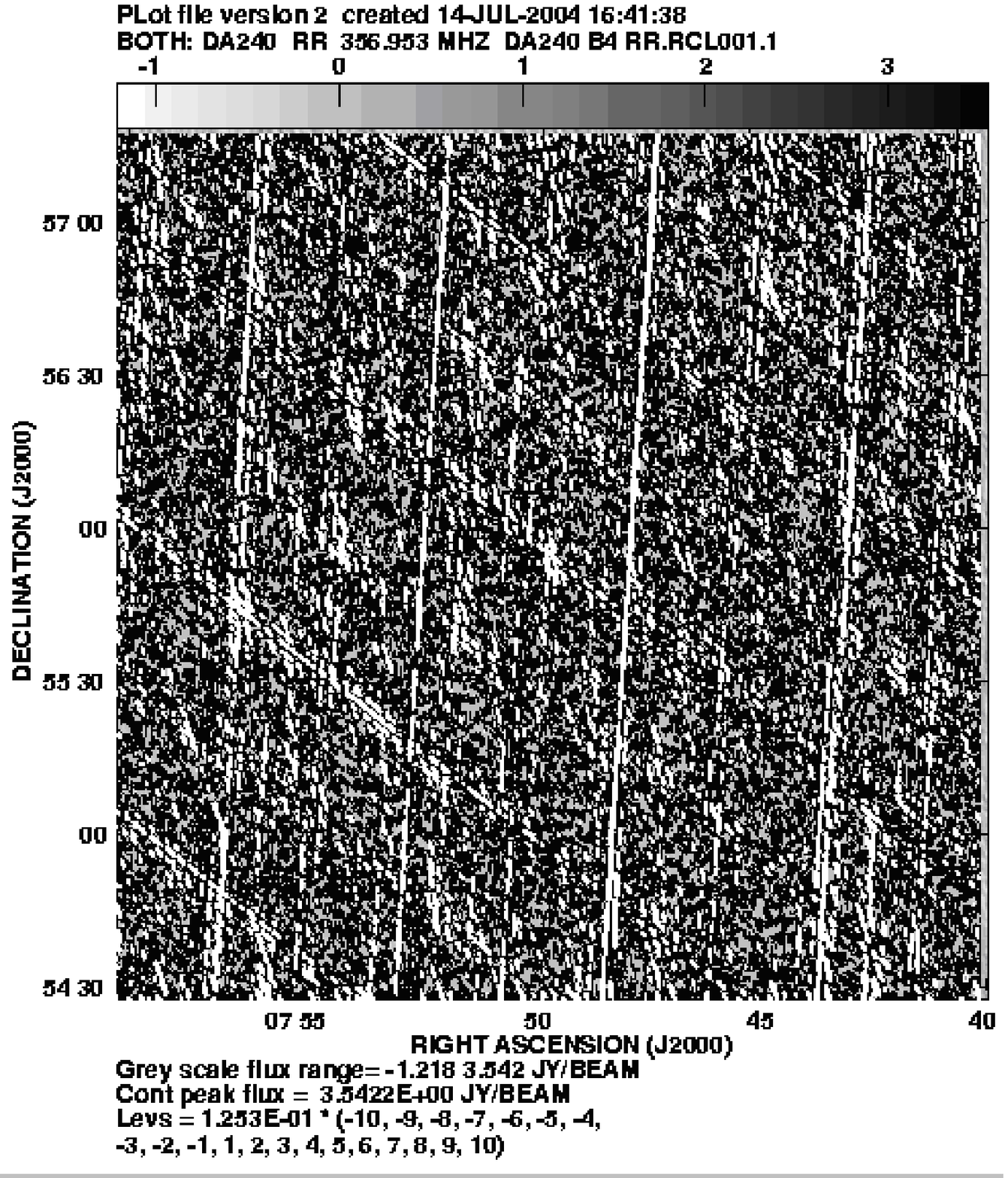}
\includegraphics[width=9.0cm,height=9.0cm]{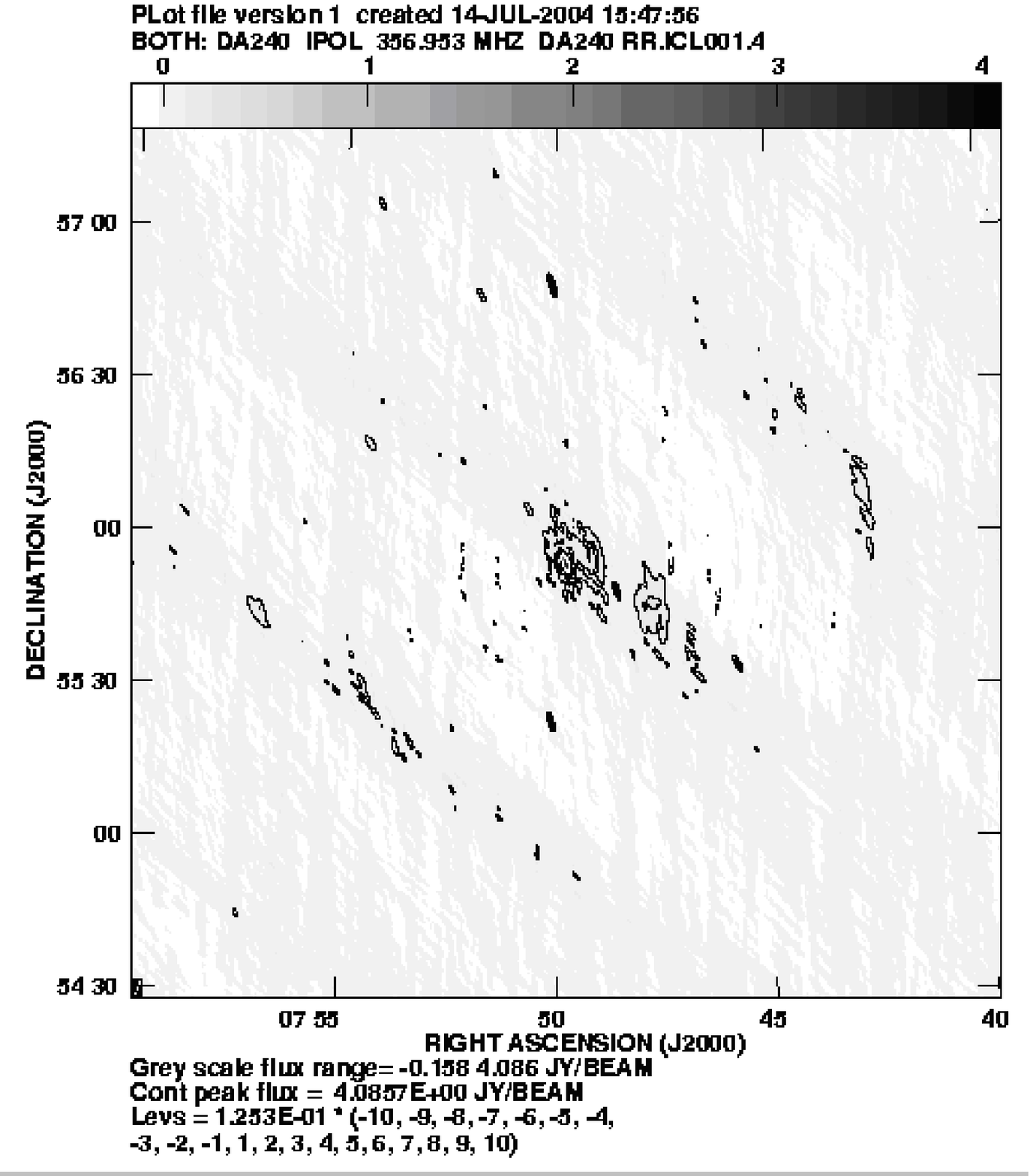}
\end{figure}

\begin{figure}
\includegraphics[width=9.0cm,height=9.0cm]{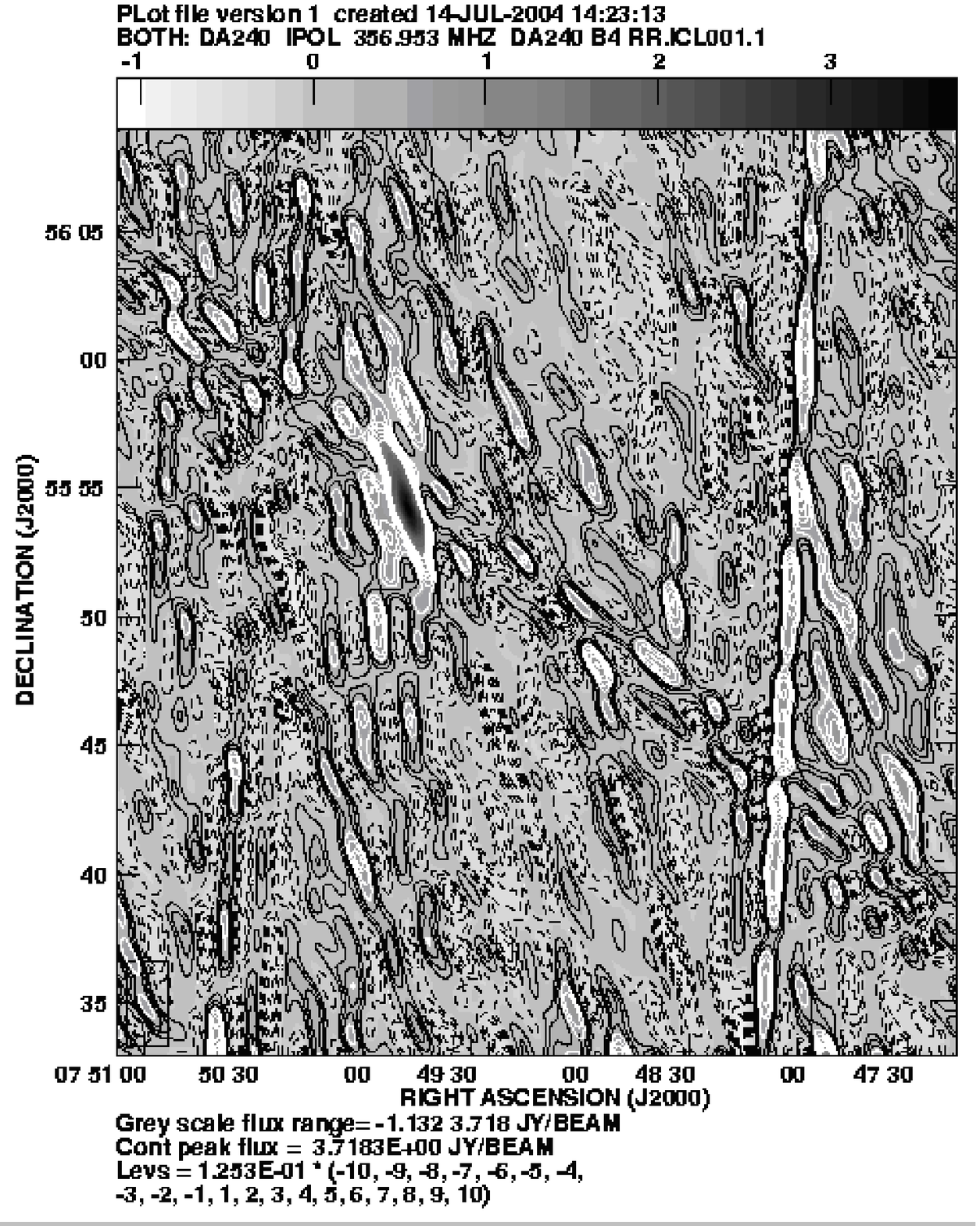}
\includegraphics[width=9.0cm,height=9.0cm]{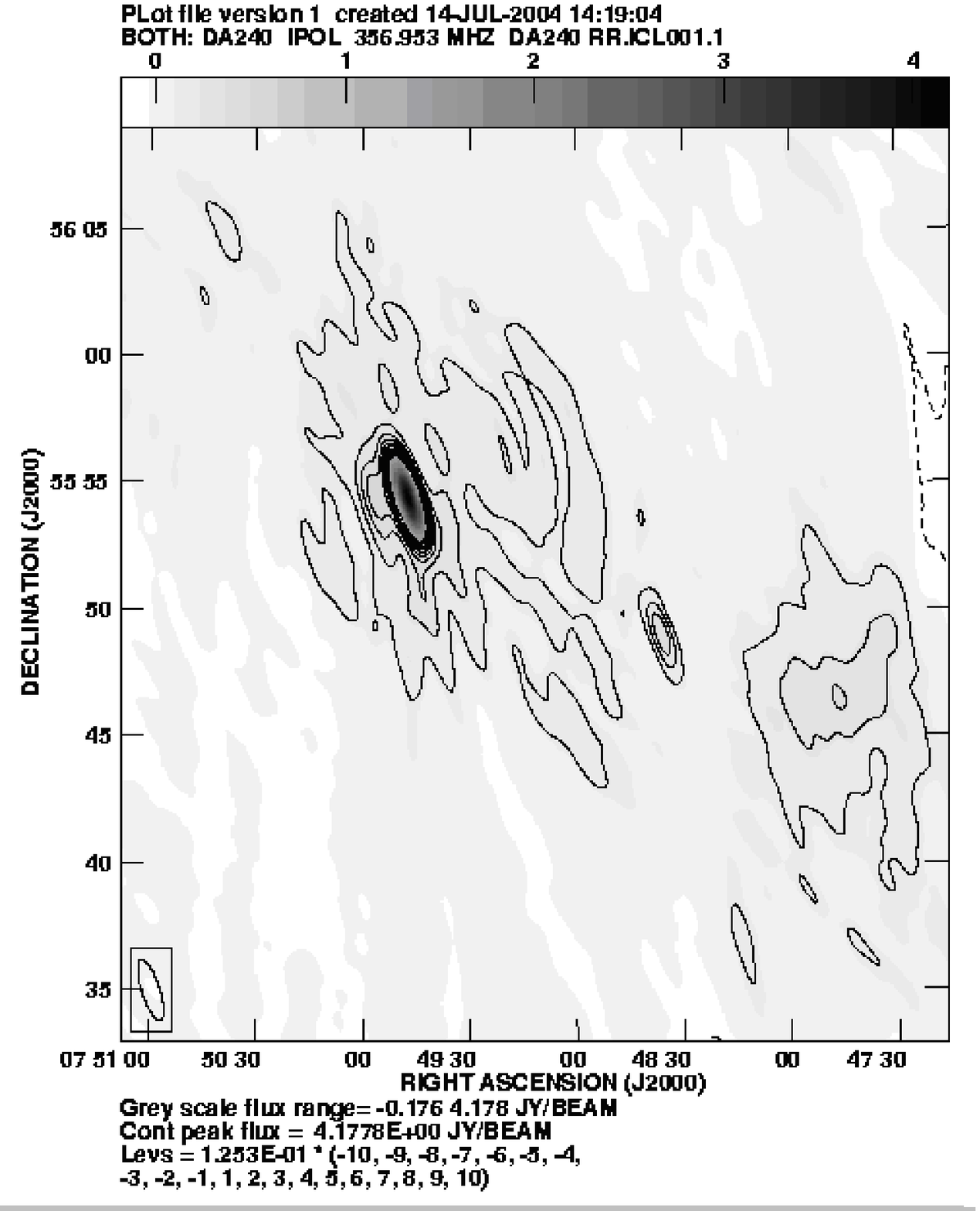}
\caption{Radio images of the source DA240 observed at WSRT, central
frequency 357 MHz, bandwidth 20 MHz, Upper row, left panel: image
without RFI mitigation; right panel: image with RFI mitigation.
Lower row:  central parts of  the  images  with and without RFI.}

\end{figure}

\clearpage

\begin{figure}
\includegraphics[width=9.0cm,height=9.0cm]{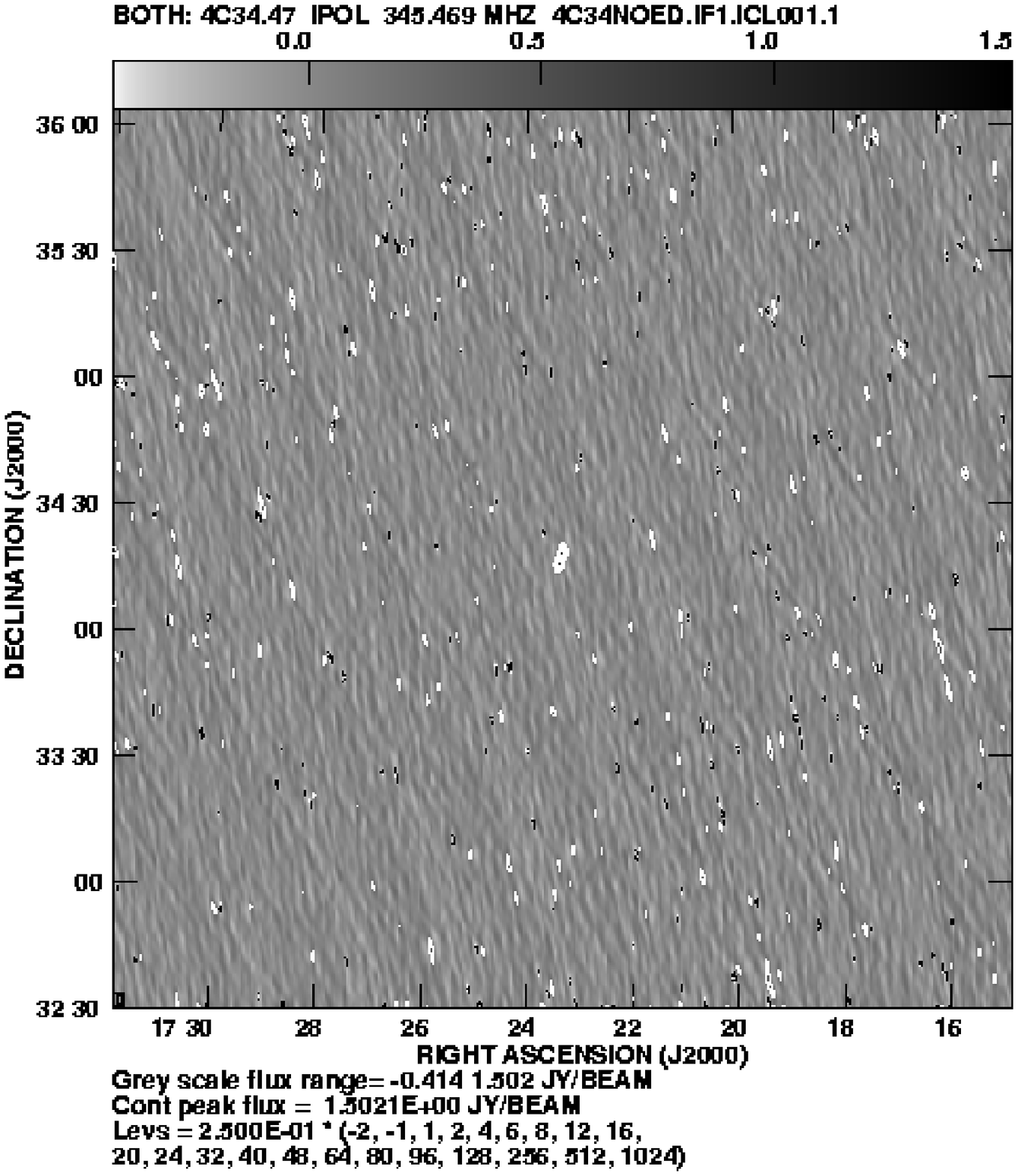}
\includegraphics[width=9.0cm,height=9.0cm]{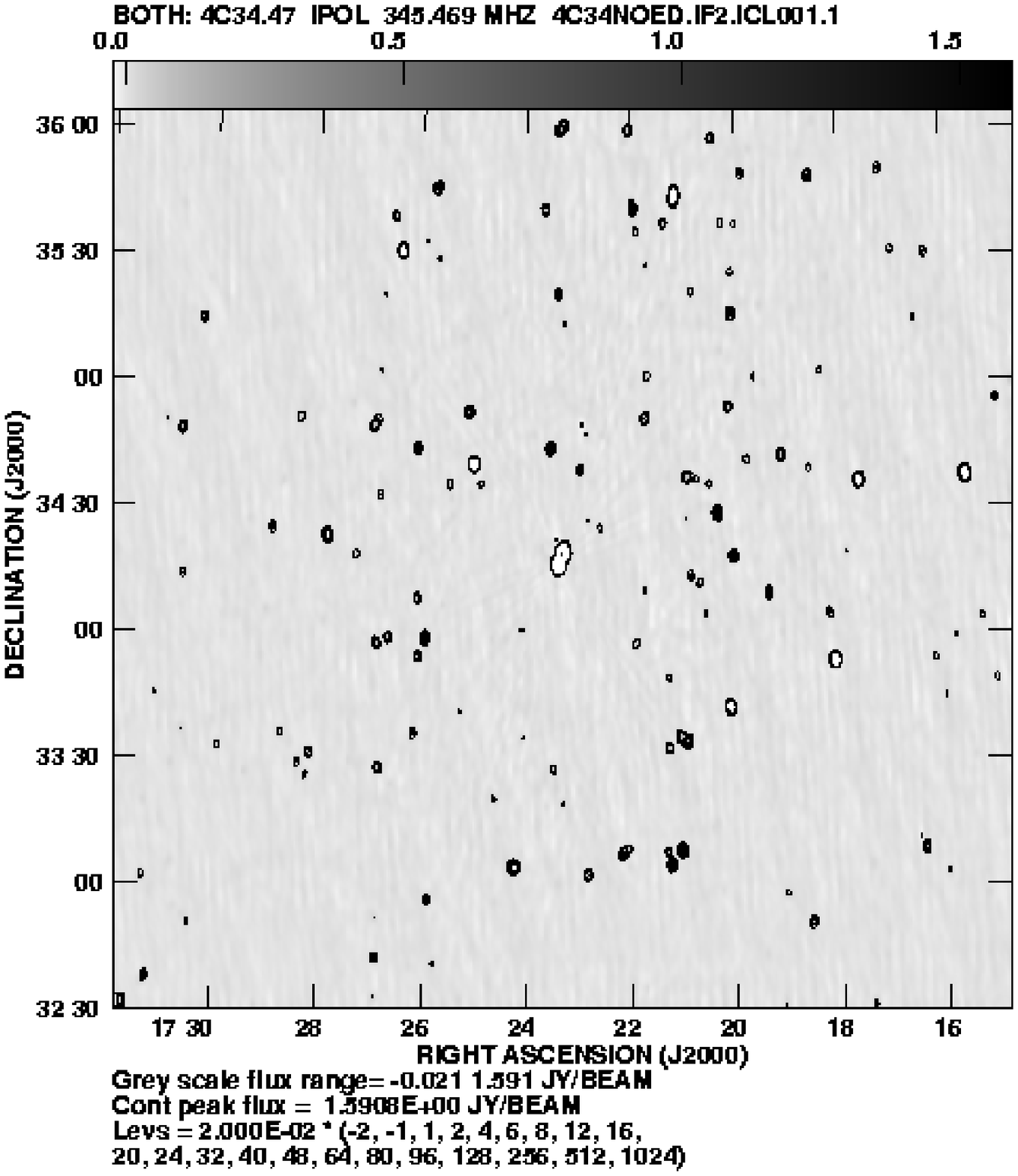}
\end{figure}

\begin{figure}
\includegraphics[width=9.0cm,height=9.0cm]{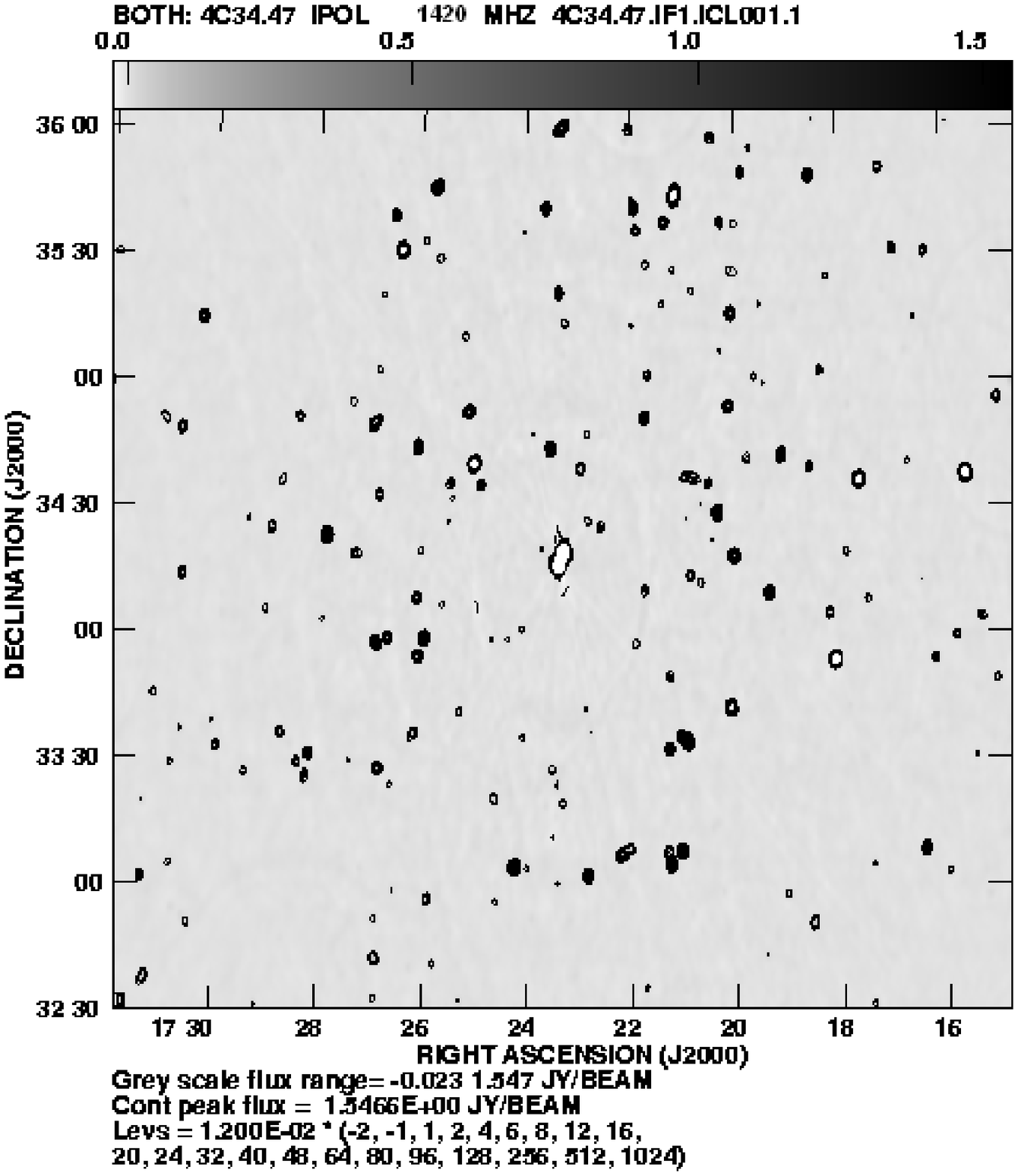}
\includegraphics[width=9.0cm,height=9.0cm]{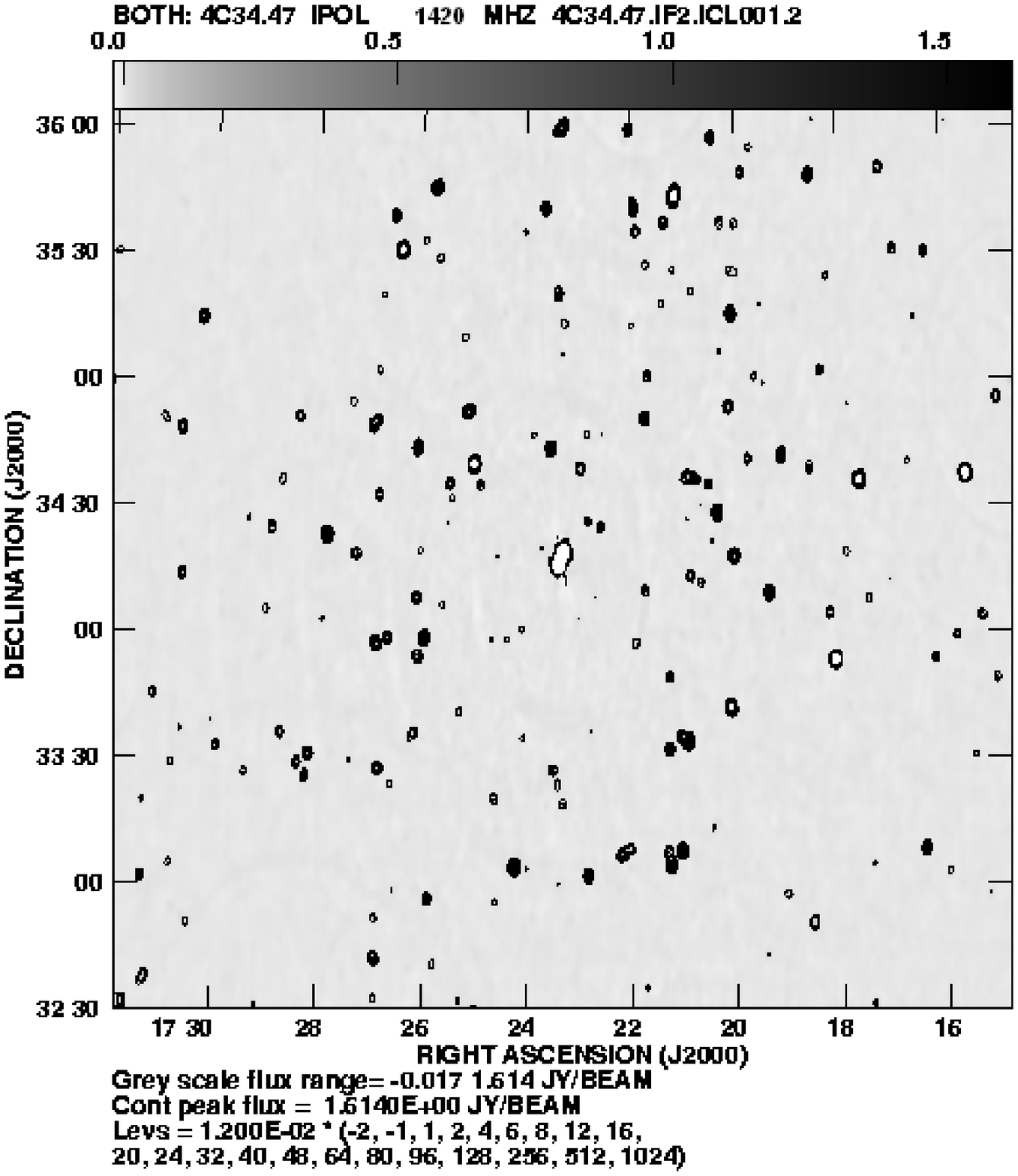}
\caption{Radio images of the source 4C34.47 observed at WSRT,
bandwidth 20 MHz.
 Upper row, left panel: central  frequency 345 MHz, without RFI mitigation system; right panel: central  frequency 345 MHz, with RFI mitigation system. Notice  the difference of intensity levels in the figures.
Lower row, left panel: central  frequency 1420 MHz without RFI
mitigation system.
 right panel: central  frequency 1420 MHz with RFI mitigation system. After subtracting one image from the other the {\it rms} noise  is  less than 0.7 mJy/beam which signifies a good similarity  in the images.}

\end{figure}

\end{document}